\documentclass[12pt]{article}

\usepackage{amsmath}
\usepackage{amssymb}
\usepackage{latexsym}
\usepackage{graphicx}

\newcommand{\pardiff}[2]{\frac{\partial#1}{\partial#2}}
\newcommand{\parddiff}[2]{\frac{\partial^2#1}{\partial#2^2}}
\newcommand{\pparddiff}[3]{\frac{\partial^2#1}{\partial#2\,\partial#3}}

\begin{document}

\nopagebreak

\title{\textbf{Hawking radiation from rotating brane black holes}\thanks{Talk given at the ``Workshop on
Dynamics and Thermodynamics of Black Holes and Naked Singularities
II'', Milan, 10-12 May 2007.}}

\author{ {\Large {Elizabeth Winstanley}}\thanks{E-mail: E. Winstanley@sheffield.ac.uk} \\[0.2in]
{\it {Department of Applied Mathematics, The University of Sheffield,}} \\
{\it {Hicks Building, Hounsfield Road, Sheffield, S3 7RH, U.K.}} }

\maketitle

\bigskip
------------------------------------------------------------------------------------------------------
\begin{abstract}
\noindent
We review recent work \cite{DHKW,CDKW,CKW} on the Hawking radiation of rotating
brane black holes, as may be produced at the LHC.
We outline the methodology for calculating the fluxes of particles, energy and angular momentum
by spin-0, spin-1/2 and spin-1 quantum fields on the brane.
We briefly review some of the key features of the emission, in particular the changes in the spectra
as the number of extra dimensions or the angular velocity of the black hole increases.
These quantities will be useful for accurate simulations of black hole events at the LHC.
\end{abstract}
\bigskip
\par------------------------------------------------------------------------------------------------------

\section{Introduction}
\label{sec:intro}

\begin{figure}
\begin{center}
\includegraphics[width=6cm,angle=270]{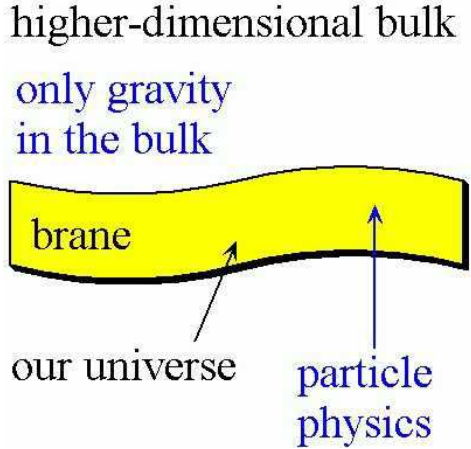}
\end{center}
\caption{Sketch of the brane-world scenario: our universe is a brane, embedded in a higher-dimensional bulk.
The standard model fields of particle physics are constrained to live on the brane; only gravity can
propagate in the bulk.}
\label{fig:brane}
\end{figure}

The formulation, nearly ten years ago, of theories with large extra dimensions \cite{ADD,RS} has
led to a highly active programme of research in gravitational physics (see also \cite{early}
for some early work in this direction).
These theories were motivated by the hierarchy problem in particle physics.
In order to resolve this problem, the fields of the standard model of particle physics are constrained
to live on a $(3+1)$-dimensional brane, which is embedded into a higher dimensional bulk
(see figure \ref{fig:brane}), so that
the total number of space-time dimensions in the bulk is $4+n$.
Only gravitational degrees of freedom are allowed to propagate in the bulk.

One consequence of these higher-dimensional theories is that the scale of quantum gravity may be very much
lower than the Planck scale $M_{P}$.
For large extra space-like dimensions with a compactification radius $L$, the fundamental scale of quantum
gravity in the bulk, $M_{*}$, is related to the Planck scale (which is now an `effective' quantum
gravity scale as seen on the brane) as follows \cite{ADD}:
\begin{equation}
M_{P}^{2} = 8 \pi M_{*}^{n+2}L^{n},
\label{eq:MP}
\end{equation}
where $n$ is the number of extra dimensions.
It is possible, via equation (\ref{eq:MP}),
to have $M_{P}\sim 10^{19}$ GeV and $M_{*}\sim 1$ TeV for various combinations
of $n$ and $L$: for example, if $n=2$ and $L\sim 1$ mm.
This lowering of the bulk scale of quantum gravity by many orders of magnitude may be expected to have
significant phenomenological implications.
In particular, particles with energy above $M_{*}$ will be able to probe quantum gravitational physics
and collisions of such particles (for example, at the LHC), may result in strong gravitational phenomena.
In this paper we focus on the formation and
subsequent evaporation of mini black holes in such trans-Planckian collisions \cite{creation},
as may occur in ground-based colliders \cite{colliders}
or ultra-high energy cosmic ray interactions in the atmosphere of the Earth
\cite{cosmic}.
We do not attempt to survey here the huge literature on these subjects, but simply
refer the reader to the reviews \cite{reviews,Kanti,Harris} for detailed discussions and
comprehensive lists of references.

The outline of this article is as follows.
We begin in section \ref{sec:BBH} by very briefly reviewing the key processes in the formation and
evaporation of brane black holes at the LHC.
Our focus is one part of the evaporation, the `spin-down' phase (which will be defined in the next section).
Our study involves quantum field theory in curved space, and the methodology is outlined in section \ref{sec:QFT}.
The key features of the  results of this study, namely the fluxes of particles,
energy and angular momentum for particles
with spin-$0$, $1$ and $1/2$, are discussed in section \ref{sec:fluxes} before we present our conclusions
in section \ref{sec:conc}.
Throughout this paper, the metric has signature $(+,-,-,-)$ and we use units in which $G=c=\hbar =k_{B}=1$.

\section{Formation and Evaporation of Brane Black Holes}
\label{sec:BBH}

\begin{figure}[h]
\begin{center}
\includegraphics[width=4cm,angle=270]{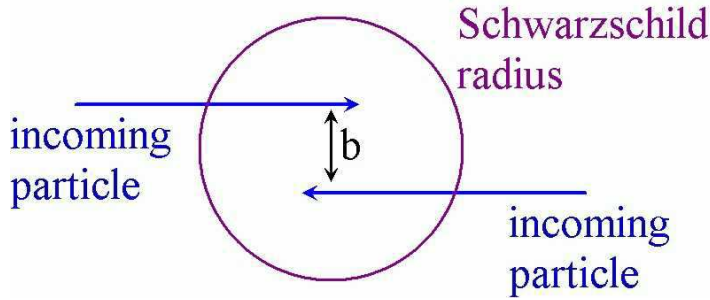}
\end{center}
\caption{Sketch of the formation of a black hole. If two particles collide in such a way that their impact
factor $b$ is smaller than the Schwarzschild radius corresponding to their centre-of-mass energy,
then it is expected that a black hole will be formed.}
\label{fig:impact}
\end{figure}

The formation of a mini-black hole in a trans-Planckian collision at the LHC may be understood heuristically
as follows (see sketch in figure \ref{fig:impact}).
Suppose we have two colliding particles with centre-of-mass energy $E$.
If their impact parameter $b$ is smaller than the Schwarzschild radius of a black hole having mass $E$,
then, classically, one would expect a black hole to be formed \cite{hoop}.
If the bulk scale of quantum gravity is as low as a few TeV, then the Schwarzschild radius of a
higher-dimensional black hole is of order $M_{*}^{-1}\ll L$,
which is very much larger than would be anticipated in general relativity.
This means that it is feasible that colliding particles at the LHC may have impact parameters smaller
than the corresponding Schwarzschild radius.
The black holes formed by this process will be higher-dimensional, extending into the bulk.
Of course, this simple argument does not accurately portray the whole process, which has been
extensively studied \cite{creation,colliders,reviews}.
Various different calculations have been done of the rate of black hole production at the LHC,
with controversial results, but including the exciting possibility of many black holes per day being formed
\cite{creation,colliders}.

\begin{figure}[h]
\begin{center}
(a)
\includegraphics[width=4cm,angle=270]{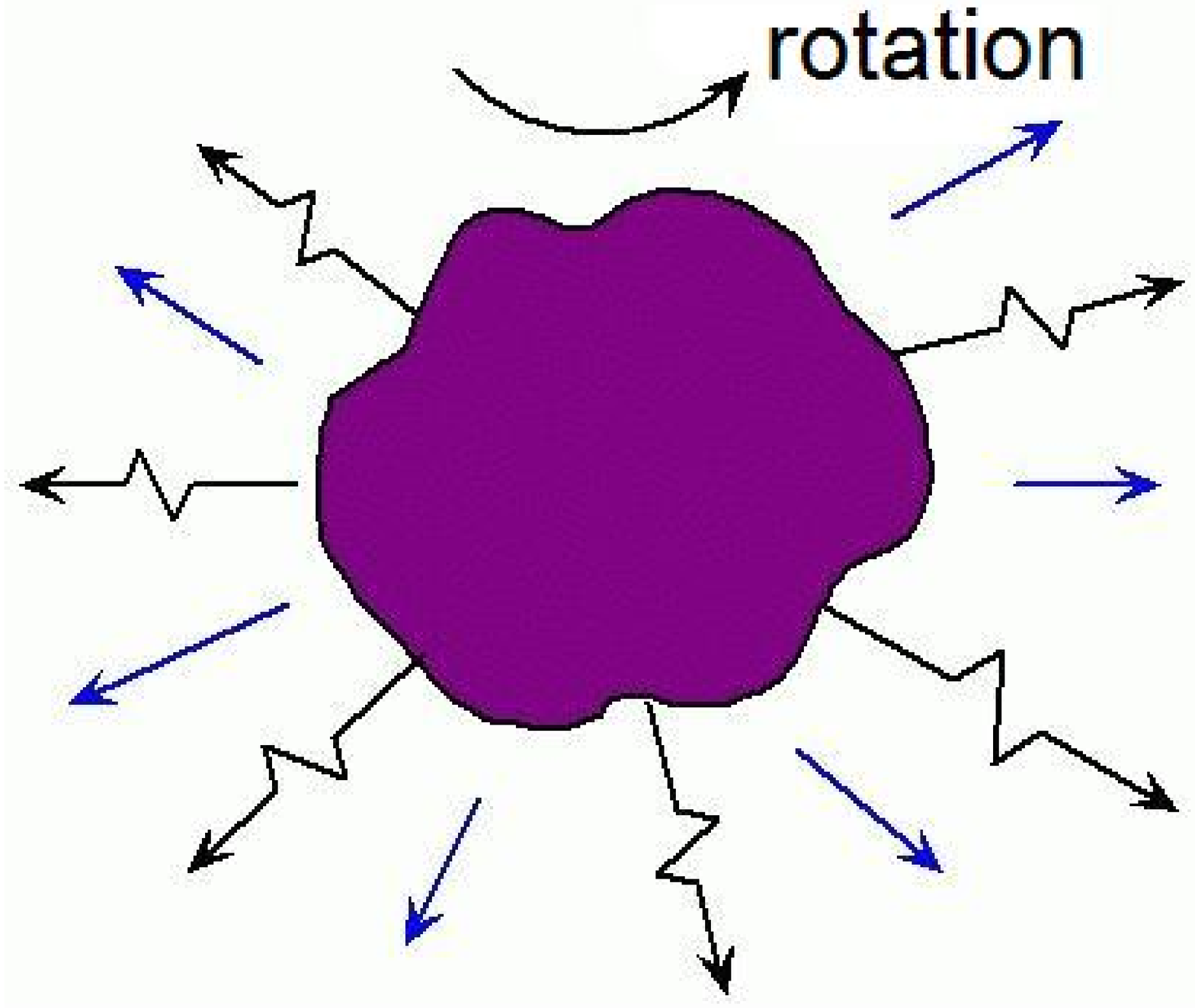}
(b)
\includegraphics[width=4cm,angle=270]{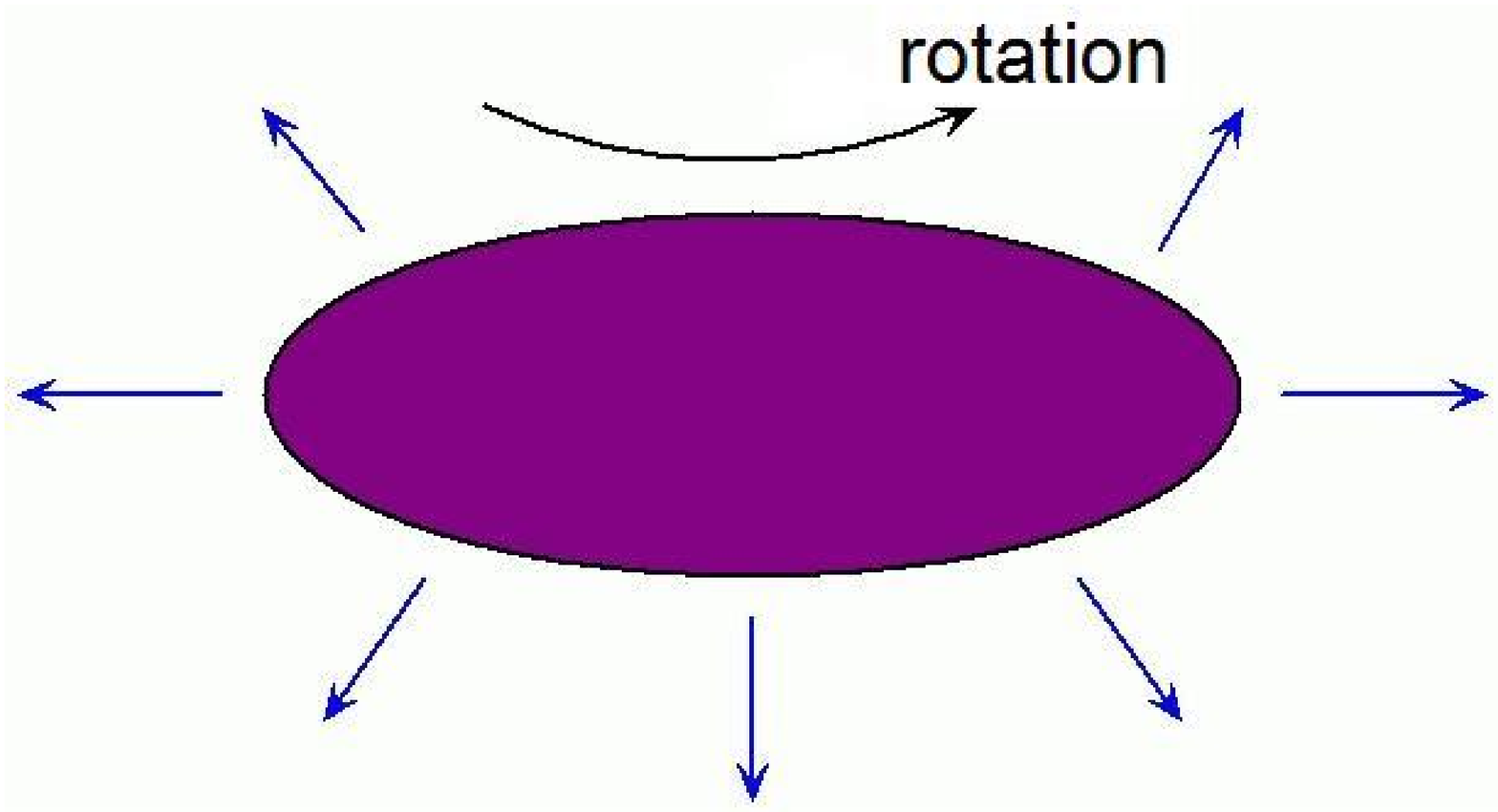} \\
(c)
\includegraphics[width=4cm,angle=270]{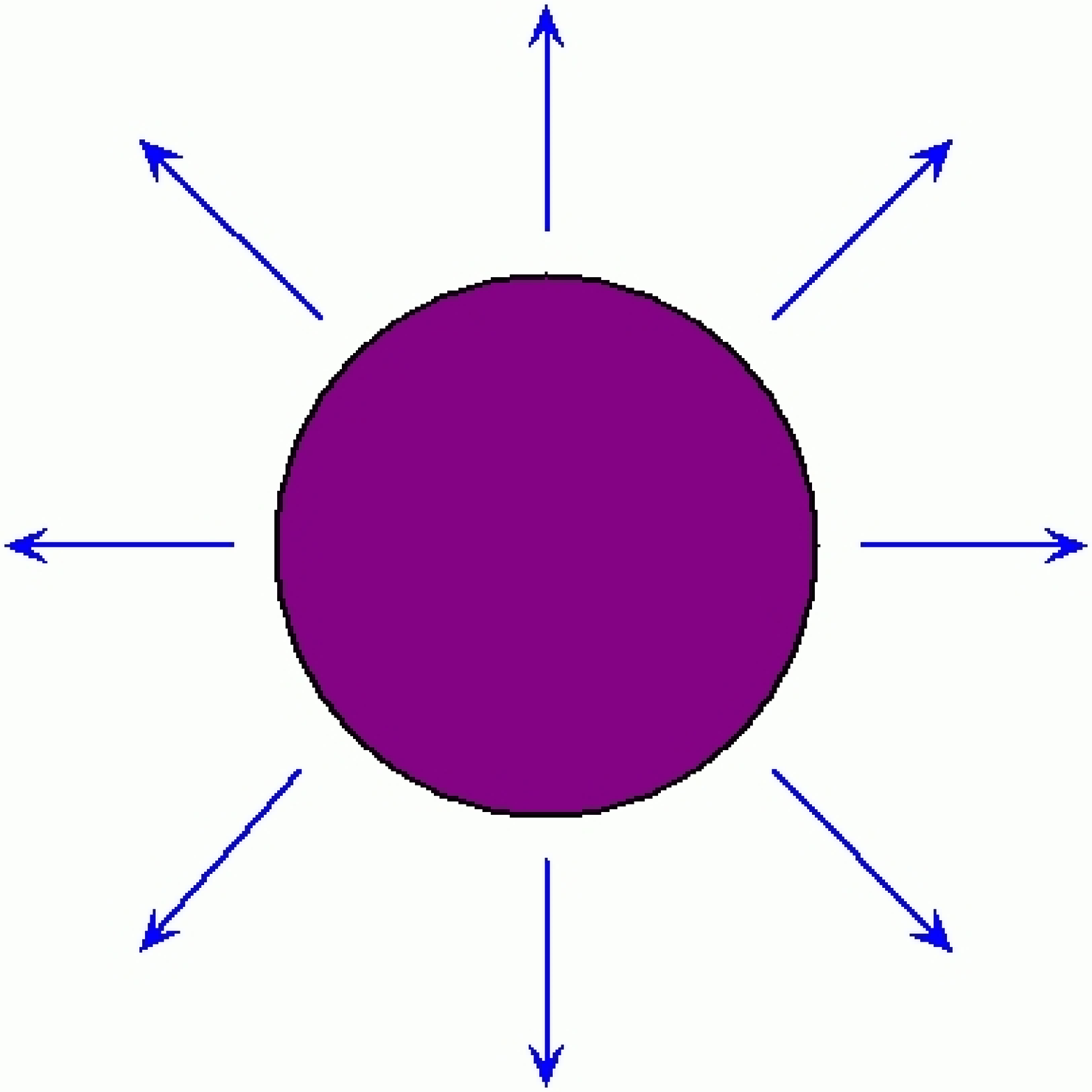}
\end{center}
\caption{Sketch of three stages of the evaporation of the black hole:
(a) the `balding' phase, where the black hole loses its asymmetries and any gauge field hair;
(b) the `spin-down' phase, where the black hole sheds it angular momentum and some mass;
(c) the `Schwarzschild' phase, where the black hole is spherically symmetric and loses mass.
The last unknown stage, the `Planck' phase, is not sketched. }
\label{fig:evap}
\end{figure}

In this article we consider black holes with masses many times greater than the bulk quantum-gravity scale $M_{*}$.
Such black holes can be considered as semi-classical objects, and quantum gravity effects can be effectively
ignored.
The main observable feature of the subsequent evolution of the black hole will be its Hawking radiation \cite{Hawking}.
This proceeds in four stages (the first three of which are sketched in figure \ref{fig:evap}).
\begin{enumerate}
\item
When the black hole forms, it will be a complicated object, highly asymmetric and endowed with non-abelian gauge
field hair from the incoming particles.
The first stage of its evolution is the {\em {`balding' phase}}, during which the gauge field hair is
shed and the asymmetries removed by gravitational radiation.
It is expected that this phase of the evolution will be extremely rapid.
\item
At the end of the `balding' phase, the black hole has no hair and will be axisymmetric, rotating extremely
rapidly.
During the {\em {`spin-down' phase}},
the black hole then evaporates through Hawking \cite{Hawking} and Unruh-Starobinskii \cite{US} radiation,
losing mostly angular momentum as well as mass.
\item
At the end of the `spin-down' phase, the black hole has shed all its angular momentum and is spherically
symmetric.
Since the black hole is Schwarzschild-like, the next phase is the {\em {`Schwarzschild' phase}}, when
the black hole continues to evaporate via Hawking radiation, but now in a spherical manner.
\item
The black hole can continue to be described semi-classically until its mass is of the order of the (higher-dimensional)
quantum gravity scale, in which case unknown quantum-gravity effects become important.
It is thought that this final stage, the {\em {`Planck' phase}} will be short-lived compared to the
`spin-down' and `Schwarzschild' phases.
\end{enumerate}
Of these four stages, the `Schwarzschild' phase is by far the simplest to analyze and correspondingly the
most work exists in the literature on this part of the evolution \cite{schwarzschild}.
Both analytic and numerical studies have been performed, for standard model fields on the brane, and scalar
and graviton fields both on the brane and in the bulk.
References \cite{schwarzschild} also include work on different types of spherically symmetric black hole background
(including black hole charge, a cosmological constant, and higher-derivative gravity) and on the emission of massive
as well as massless quantum fields.

Our focus in the present article is the `spin-down' phase.
After the extensive work on the `Schwarzschild' phase, the `spin-down' phase has recently been the focus
of much attention in the literature \cite{DHKW,CDKW,CKW,spin}.
In this article we concentrate on the emission of quantum fields of spin $0$, $1/2$ and $1$ on the brane,
although there is work in literature \cite{spin} which includes emission into the bulk.

\section{Quantum Fields on a Rotating Brane Black Hole}
\label{sec:QFT}

\begin{figure}[h]
\begin{center}
\includegraphics[width=5cm,angle=270]{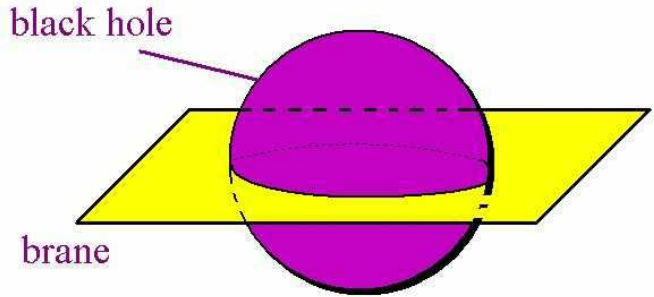}
\end{center}
\caption{Sketch of the geometry of the higher-dimensional black hole. The black hole
extends into the bulk, and we consider the `slice' of the black hole which intersects with the brane
(our universe).}
\label{fig:geometry}
\end{figure}

If we assume that the colliding particles which form the black hole are constrained to lie on
an infinitely thin brane, so that they have a non-zero impact parameter only in the brane, then the
black hole thus produced will have its plane of rotation in the brane.
During the spin-down phase, the higher-dimensional black hole geometry will therefore be of the
Myers-Perry \cite{MP} form, with a single non-zero angular momentum parameter $a$:
\begin{eqnarray}
ds^2 & = & \left( 1-\frac{\mu}{\Sigma\,r^{n-1}}\right) dt^2 +
\frac{2 a \mu \sin^2\theta}{\Sigma\,r^{n-1}}\,dt \, d\varphi
-\frac{\Sigma}{\Delta}\,dr^2 -\Sigma\,d\theta^2
\nonumber \\
& & - \left(r^2+a^2+\frac{a^2 \mu \sin^2\theta}{\Sigma\,r^{n-1}}\right)
\sin^2\theta\,d\varphi^2 - r^2 \cos^2\theta\, d\Omega_{n},
\label{eq:rot-metric}
\end{eqnarray}
where
\begin{equation}
\Delta = r^2 + a^2 -\frac{\mu}{r^{n-1}}\,, \qquad
\Sigma=r^2 +a^2\,\cos^2\theta\,,
\label{Delta}
\end{equation}
and $d\Omega_n$ is the line-element on a unit $n$-sphere.
In this paper we are interested in the emission of particles on the brane, and so we first need
to determine the effective, four-dimensional, black hole metric on which the quantum fields propagate.
We do this by fixing the values of the additional angular co-ordinates used in (\ref{eq:rot-metric})
to describe the extra dimensions.
Effectively we are taking a `slice' through the higher-dimensional black hole geometry, as sketched in
figure \ref{fig:geometry}.
The four-dimensional metric resulting from this process is:
\begin{eqnarray}
ds^2 & = & \left( 1-\frac{\mu}{\Sigma\,r^{n-1}}\right) dt^2 +
\frac{2 a \mu \sin^2\theta}{\Sigma\,r^{n-1}}\,dt \, d\varphi
-\frac{\Sigma}{\Delta}\,dr^2 -\Sigma\,d\theta^2
\nonumber \\
& &   - \left(r^2+a^2+\frac{a^2 \mu \sin^2\theta}{\Sigma\,r^{n-1}}\right)
\sin^2\theta\,d\varphi^2 .
\label{eq:induced}
\end{eqnarray}

The location of the black hole event horizon is given by solving the equation $\Delta =0$, which, for
$n> 1$, gives the unique solution
\begin{equation}
r_{h}^{n+1}=\frac {\mu}{(1+a_*^2)},
\end{equation}
where $a_{*}=a/r_{h}$.
For $n=0$, $1$, there is a maximum possible value of $a$ which gives a real solution to the equation $\Delta = 0$,
but for $n>1$ there is no such {\em {a priori}} bound on $a$.
However, there is an upper bound on $a$ from the requirement that the black hole be formed by the collision of
two particles, as shown in figure \ref{fig:impact}, namely \cite{Harris}
\begin{equation}
a_{*}^{\rm {max}} = \frac {1}{2} \left( n +2 \right ) .
\end{equation}

We wish to study quantum fields of spin $0$, $1/2$ and $1$ on the background (\ref{eq:induced}).
To do this, we employ the Newman-Penrose formalism \cite{NP}, and follow the approach of Teukolsky \cite{Teukolsky}.
As in Teukolsky's analysis of the perturbations of Kerr black holes, here we find that the
wave equations can be written as one single `master' equation, with the `helicity' $h=(+s,-s)$
of the field as a parameter in the equation \cite{Kanti,Teukolsky}:
\begin{eqnarray}
\Sigma\cdot T_{h} & = &
\left[\frac{(r^2+a^2)^2}{\Delta}-a^2\sin^2\theta\right]\parddiff{\Omega_{h}}{t}+
\frac{2a(r^2+a^2-\Delta)}{\Delta}\pparddiff{\Omega_{h}}{t}{\phi}+
\left[\frac{a^2}{\Delta}-\frac{1}{\sin^2\theta}\right]\parddiff{\Omega_{h}}{\phi} ;
\nonumber \\ & &
-\Delta^{-h}\frac{\partial}{\partial r}\left(\Delta^{h+1}\pardiff{\Omega_{h}}{r}\right)-
\frac{1}{\sin\theta}\frac{\partial}{\partial \theta}\left(\sin\theta\pardiff{\Omega_{h}}{\theta}\right)-
2h\left[\frac{a\Delta'(r)}{2\Delta}+\frac{i\cos\theta}{\sin^2\theta}\right]\pardiff{\Omega_{h}}{\phi}
\nonumber \\ & &
+2h\left[\frac{2\Delta r-(r^2+a^2)\Delta'(r)}{2\Delta}+r+ia\cos\theta\right]\pardiff{\Omega_{h}}{t}
\nonumber \\ & &
+ \left[h^2\cot^2\theta-h+h(2-\Delta''(r))\delta_{h,|h|}\right]\Omega_{h};
\label{eq:Teuk.eq.}
\end{eqnarray}
where $\Omega _{h}=\Omega _{h}(t,r,\theta ,\phi)$ represents the spin-field
perturbation and $T_{h}$ the source term.
The reader is referred to \cite{Kanti,Teukolsky} for the details of this derivation and the precise
definitions of $\Omega _{h}$ and $T_{h}$ for each spin.
The `master' equation (\ref{eq:Teuk.eq.}) is separable for any value of the helicity $h$,
so that its solution can be
written as a sum over the Fourier modes:
\begin{eqnarray}
\Omega_{h}(t,r,\theta,\phi)&= &\int_{-\infty}^{+\infty}d\omega
\sum_{\ell=|h|}^{+\infty}\sum_{m=-\ell}^{+\ell}{}_{h}a_{\Lambda}\ {}_{h}\Omega_{\Lambda}(t,r,\theta,\phi) ,
\nonumber \\
{}_{h}\Omega_{\Lambda}(t,r,\theta,\phi)&=&
{}_{h}R_{\Lambda}(r){}_{h}S_{\Lambda}(\theta)e^{-i\omega t}e^{+im\phi} ,
\label{eq:Fourier expansion for Omega_h}
\end{eqnarray}
where ${}_{h}a_{\Lambda}$ are the Fourier coefficients and
the set of `quantum' numbers is given by $\Lambda\equiv \{\ell m\omega\}$.
The radial and angular ODEs resulting from the separation of variables of the `master' equation (\ref{eq:Teuk.eq.})
are, respectively:
\begin{equation}
\label{eq:radial teuk. eq.}
0= \Delta^{-h }\frac{d}{dr}\left(\Delta^{h+1}\frac{d_{h}R_{\Lambda}}{dr}\right)+
\left[\frac{K^2-ihK\Delta'(r)}{\Delta}+4ih \omega r+h(\Delta''(r)-2)\delta_{h,|h|}-{}_{h}\lambda_{\Lambda}\right]
{}_{h}R_{\Lambda};
\end{equation}
and
\begin{eqnarray}
0 & = &
\left[
\frac{d}{dx}\left((1-x^{2})\frac{d}{dx}\right)+(a\omega)^{2}(x^{2}-1)-2h a\omega x
\right. \nonumber \\ & & \left.
-\frac{(m+hx)^{2}}{1-x^{2}}+{}_{h}\lambda_{\Lambda}+2ma\omega+h
\right] {}_{h}S_{\Lambda}(x) ;
\label{eq:ang. teuk. eq.}
\end{eqnarray}
where $x\equiv \cos \theta$ and ${}_{h}\lambda_{\Lambda}$ is the constant of separation between
the angular and radial equations.

The angular equation (\ref{eq:ang. teuk. eq.}) has solutions which are the {\em {spin-weighted spheroidal
harmonics}} \cite{spheroidal}.
Properties of these special functions and numerical methods for finding them can be found in the literature
\cite{DHKW,CDKW,CKW,spheroidal}.
The spin-weighted spheroidal harmonics ${}_{h}S_{\Lambda }$ are normalized according to
\begin{equation}
\int _{0}^{\pi } d\theta \, \sin \theta \, {}_{h}S_{\Lambda }^{2} (\cos \theta ) = 1.
\end{equation}
The angular equation (\ref{eq:ang. teuk. eq.}) has to be solved first because it is necessary to determine
the separation constant ${}_{h}\lambda_{\Lambda}$ which also appears in the radial equation (\ref{eq:radial teuk. eq.}).
Having found the separation constant, we then numerically integrate the radial equation (\ref{eq:radial teuk. eq.}).
There are some subtleties about how this is done in practice, depending on the dimensionality of space-time
and the spin of the quantum field.
We do not discuss these issues here; further details of our numerical methods can be found in \cite{DHKW,CDKW,CKW}.
We solve the radial equation (\ref{eq:radial teuk. eq.}) with the boundary conditions that the solutions
must be ingoing at the future event horizon (this corresponds to the definition of the usual `in' modes).
From our solution of the radial equation, a key quantity, which will be essential later in our analysis, is the
transmission coefficient ${\mathbb {T}}_{\Lambda }$, which is defined as the ratio of energy fluxes:
\begin{equation}
\mathbb{T}_{\Lambda } =
\frac{d E^{\text{(tra)}}_{\Lambda }/dt}{d E^{\text{(inc)}}_{\Lambda}/dt} ,
\label{eq:transcoeff}
\end{equation}
with $dE^{\text{(inc)}}/dt$ and $dE^{\text {(tra)}}/dt$ denoting, respectively, the incident
energy flux on the black hole and the energy flux transmitted down the event horizon, for
a particular field mode with quantum numbers $\Lambda $.

We are interested in the fluxes of particles, energy and angular momentum from the evaporating black hole due
to Hawking radiation, the Hawking temperature of the black hole (\ref{eq:induced}) being
\begin{equation}
T_\text{H}=\frac{(n+1)+(n-1)a_*^2}{4\pi(1+a_*^2)r_{h}} .
\label{temperature}
\end{equation}
The evaporating black hole is described by the `past' Unruh vacuum $| U^{-} \rangle $ \cite{Unruh}
on the metric (\ref{eq:induced}).
The fluxes of energy $E$ and angular momentum $J$ can be expressed in terms of components of the renormalized
stress-energy tensor:
\begin{eqnarray}
\frac {dE}{dt} & = &
\int _{S_{\infty }} \langle U^{-} |T^{rt} |U^{-} \rangle _{\rm {ren}}
\, r^{2} \sin \theta \, d\theta \, d\varphi ;
\nonumber \\
\frac {dJ}{dt} & = &
\int _{S_{\infty }} \langle U^{-} | T^{r}_{\varphi } | U^{-} \rangle _{\rm {ren}}
\, r^{2} \sin \theta \, d\theta \, d\varphi ;
\label{eq:RSET}
\end{eqnarray}
while the particle fluxes are given by the expectation values of the appropriate particle number operator.
Fortunately the stress-energy tensor components in (\ref{eq:RSET}) do not require renormalization
(for spin 0, the argument is presented in \cite{Unruh}, and can be generalized to higher spins, using
the renormalization counterterms presented in \cite{Christensen}).

In order to construct the state $|U^{-} \rangle $, we start by defining the usual `in' and `up' modes
for the quantum field \cite{Unruh}.
We then take a basis of field modes for which the `in' modes have positive frequency with respect to co-ordinate
time $t$ near infinity, while the `up' modes have positive frequency with respect to Kruskal time near the
past event horizon.
We do not describe the details here; instead we refer the reader to the relevant literature for the cases of
spin 0 \cite{DHKW,Unruh,spin0}, spin 1/2 \cite{CDKW,spin05} and spin 1 \cite{CKW,spin1}.
After computing the expectation value of the stress-energy tensor in the state
$\left| U^{-} \right\rangle $, the total fluxes of particles, energy and angular
momentum emitted by the black hole per unit time and frequency,
for each species, are given by \cite{DHKW,CDKW,CKW}:
\begin{eqnarray}
\frac {d^{2}N}{dt  d\omega}  & = &
\frac {\left( 1 + \delta _{s,1} \right) }{2\pi} \sum _{\ell =s}^{\infty }
\sum _{m=-\ell }^{\ell }
\frac {1}{\exp\left(\tilde{\omega}/T_\text{H}\right) \pm 1}
{\mathbb {T}}_{\Lambda } ;
\label{eq:flux} \\
\frac {d^{2}E}{dt  d\omega }  &  = &
\frac {\left( 1 + \delta _{s,1} \right) }{2\pi} \sum _{\ell =s}^{\infty }
\sum _{m=-\ell }^{\ell }
\frac {\omega }{\exp\left(\tilde{\omega}/T_\text{H}\right) \pm 1}
{\mathbb {T}}_{\Lambda } ;
\label{eq:power} \\
\frac {d^{2}J}{dt  d\omega}  & = &
\frac {\left( 1 + \delta _{s,1} \right) }{2\pi} \sum _{\ell =s}^{\infty }
\sum _{m=-\ell }^{\ell }
\frac {m}{\exp\left(\tilde{\omega}/T_\text{H}\right) \pm 1}
{\mathbb {T}}_{\Lambda } ;
\label{eq:angmom}
\end{eqnarray}
where
\begin{equation}
{\tilde {\omega }}= \omega - m\Omega _{H},
\end{equation}
and $\Omega _{H}$ is the angular velocity of the event horizon, given by
\begin{equation}
\Omega _{H} = \frac {a_{*}}{\left( 1 + a_{*}^{2} \right) r_{h}} .
\end{equation}
In equations (\ref{eq:flux}--\ref{eq:angmom}), the positive sign in the denominator is used for fermions,
while the negative sign gives the correct Planck factor for bosons.
Note that there is an additional factor of $2$ in the gauge boson (spin-1) case, which results from a sum over
the parity of the field modes.

The emission rates (\ref{eq:flux}--\ref{eq:angmom}) describe the different
fluxes emitted by the black hole over the whole solid angle $\Omega_2^2=4 \pi$,
and follow by performing an integration over the angular coordinates
$\theta$ and $\varphi $. If we go one step backwards, we may derive the
angular distribution of the emitted radiation by displaying the exact
dependence of the differential rates on the latitudinal angle $\theta$,
and write
\begin{eqnarray}
\frac {d^{3}N}{d(\cos\theta)dt  d\omega}  & = &
\frac { \left( 1 + \delta _{s,1} \right)  }{4\pi} \sum _{\ell =s}^{\infty }
\sum _{m=-\ell }^{\ell }
\frac {1}{\exp\left(\tilde{\omega}/T_\text{H}\right) \pm 1}
{\mathbb {T}}_{\Lambda }
 \left({}_{-h}S_{\Lambda}^2+{}_{+h}S_{\Lambda}^2\right);
\label{eq:fluxang} \\
\frac {d^{3}E}{d(\cos\theta)dt  d\omega }  &  = &
\frac {\left( 1 + \delta _{s,1} \right)  }{4\pi} \sum _{\ell =s}^{\infty }
\sum _{m=-\ell }^{\ell }
\frac {\omega }{\exp\left(\tilde{\omega}/T_\text{H}\right) \pm 1}
{\mathbb {T}}_{\Lambda }
\left({}_{-h}S_{\Lambda}^2+{}_{+h}S_{\Lambda}^2\right).
\label{eq:powerang}
\end{eqnarray}
Again, there is an additional factor of $2$ in the spin-1 case, arising from a sum over the parity of the field modes.
Note that in the spin-0 (scalar) case, the two helicities $h$, $-h$ are equal to zero, so we simply have an additional
factor of two.

We will now outline the key properties of the fluxes (\ref{eq:flux}-\ref{eq:powerang}) for each of the particle species.

\pagebreak

\section{Fluxes of Particles, Energy and\\ Angular Momentum}
\label{sec:fluxes}

We now briefly outline some of the key features of our results. For further details, and a more wide-ranging discussion
(including, for example, the total emissions), we refer the reader to the literature \cite{DHKW,CDKW,CKW,spin}.
Detailed studies of the Hawking radiation from four-dimensional black holes can be found in \cite{classics}.

\begin{figure}[h]
\begin{center}
\includegraphics[width=5.5cm,angle=270]{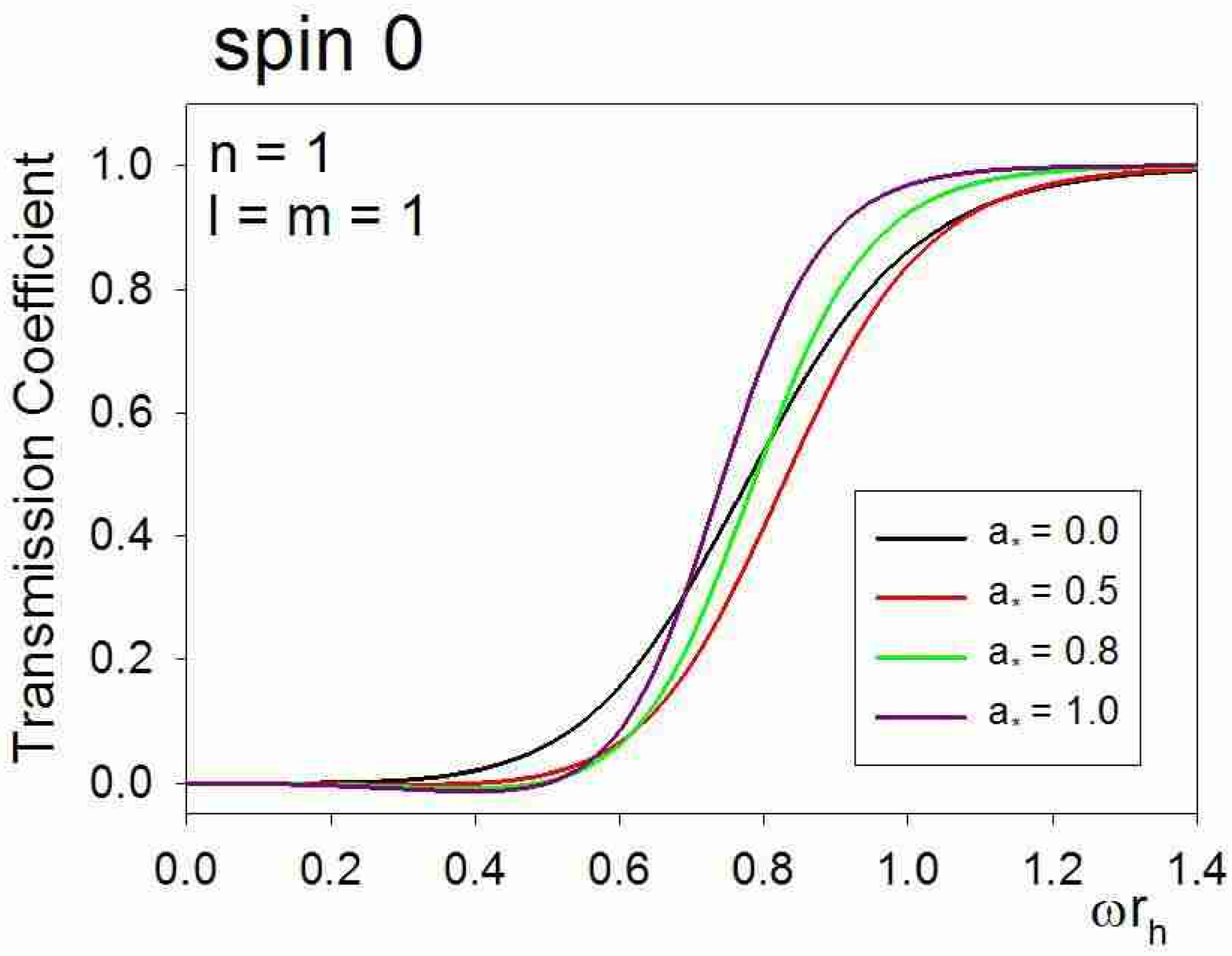}
\includegraphics[width=5.5cm,angle=270]{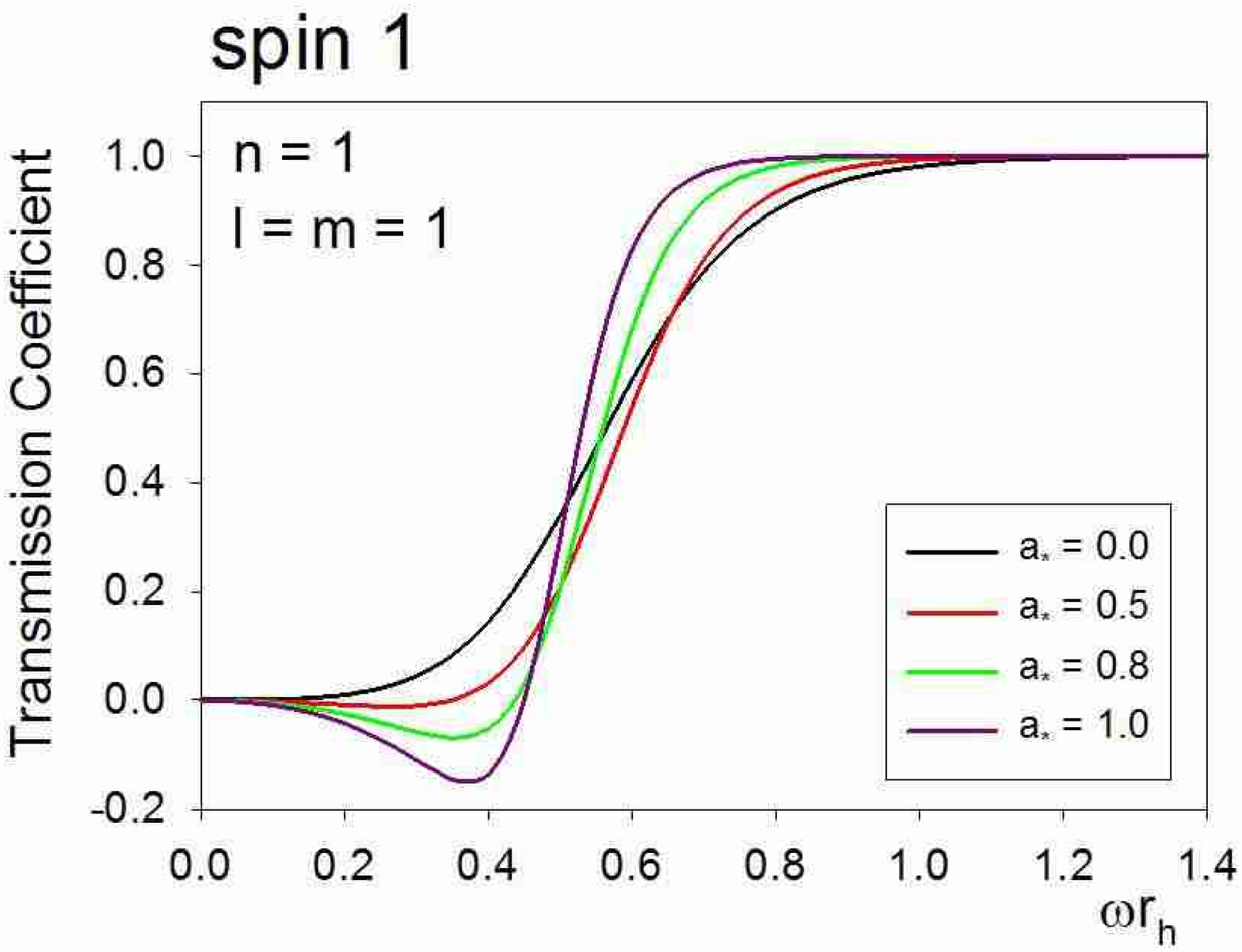}
\end{center}
\caption{Transmission coefficient ${\mathbb {T}}_{\Lambda }$ for spin-0 and spin-1 fields,
for the $\ell = 1 = m$ mode with $n=1$, for various values of $a_{*}$.}
\label{fig:transmission}
\end{figure}

The first quantity of interest is the transmission coefficient ${\mathbb {T}}_{\Lambda }$ (\ref{eq:transcoeff}).
This is plotted for the mode $\ell = 1 =m$ and $n=1$ in figure \ref{fig:transmission}, for the cases of spin-0 and spin-1
and different values of $a_{*}$.
The interesting feature of the plots in figure \ref{fig:transmission} is that in both cases the transmission coefficient
is negative for small values of $\omega r_{h}$. This effect is small for the spin-0 mode but significant for the spin-1
mode.
This is, of course, the famous super-radiance effect: an incoming wave is reflected back to infinity with amplitude
greater than it had originally.
This effect is observed only for bosons, and does not occur for spin-1/2 modes.

\begin{figure}[p]
\begin{center}
\includegraphics[width=5cm,angle=270]{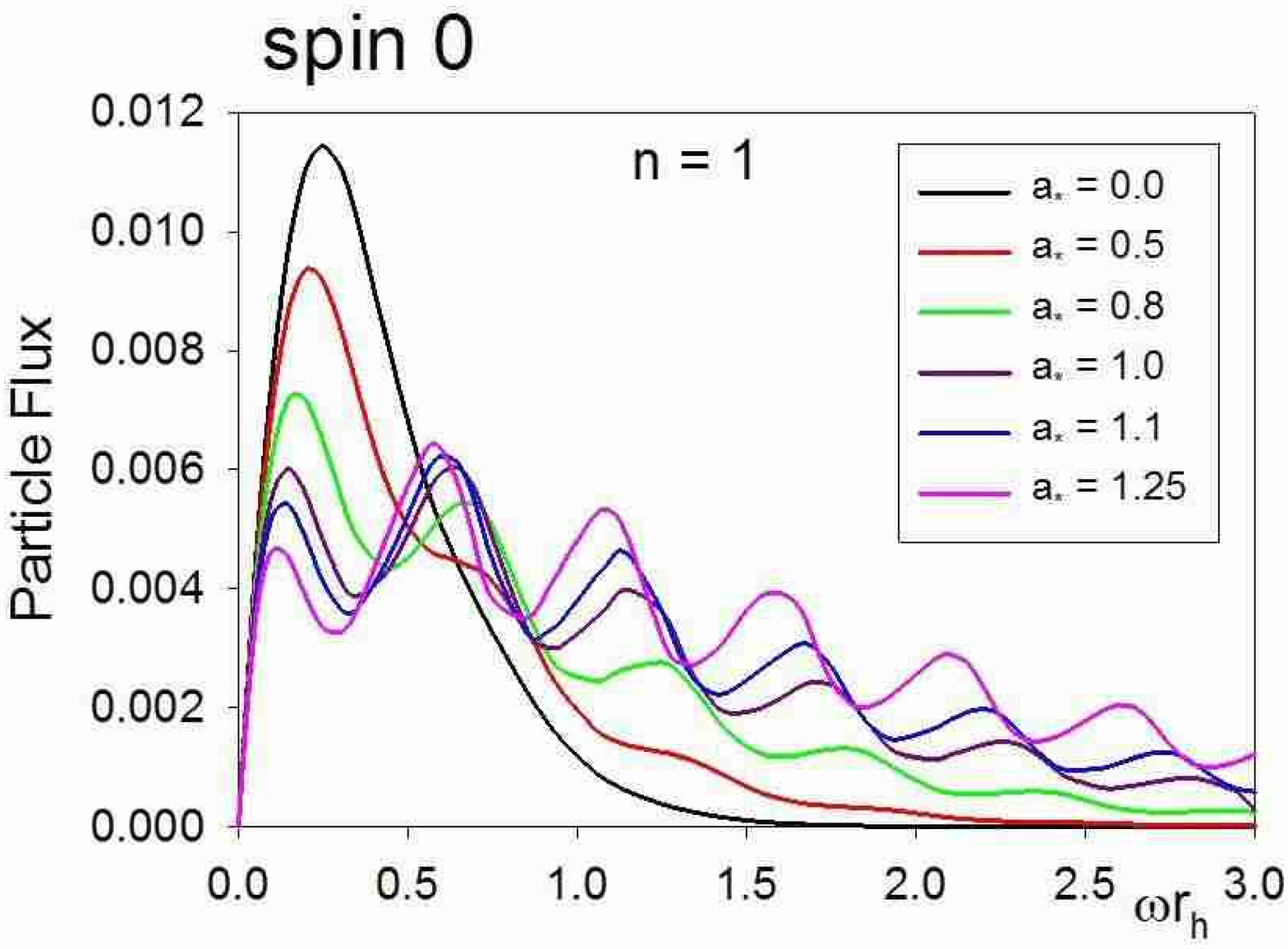}
\includegraphics[width=5cm,angle=270]{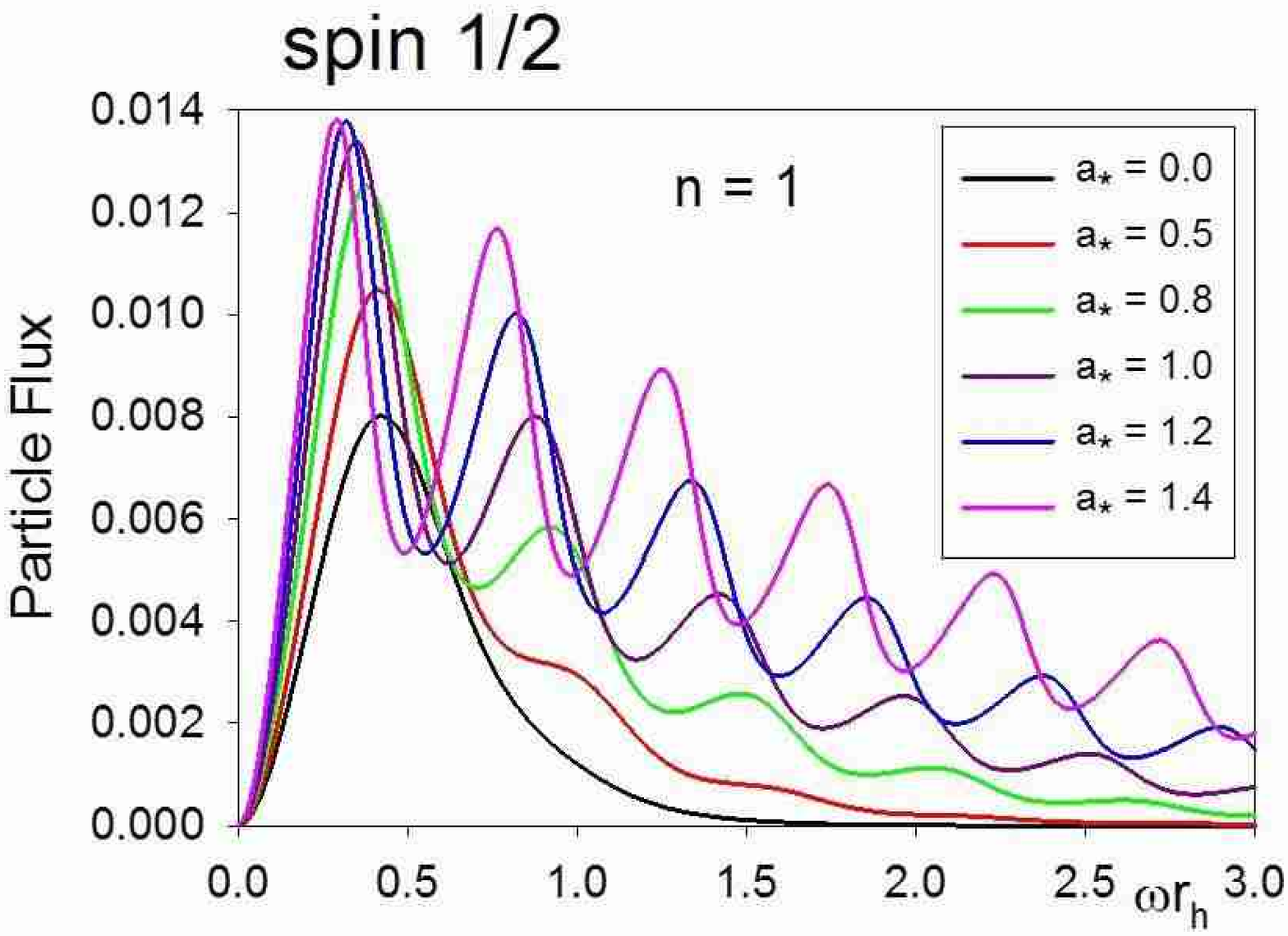}
\includegraphics[width=5cm,angle=270]{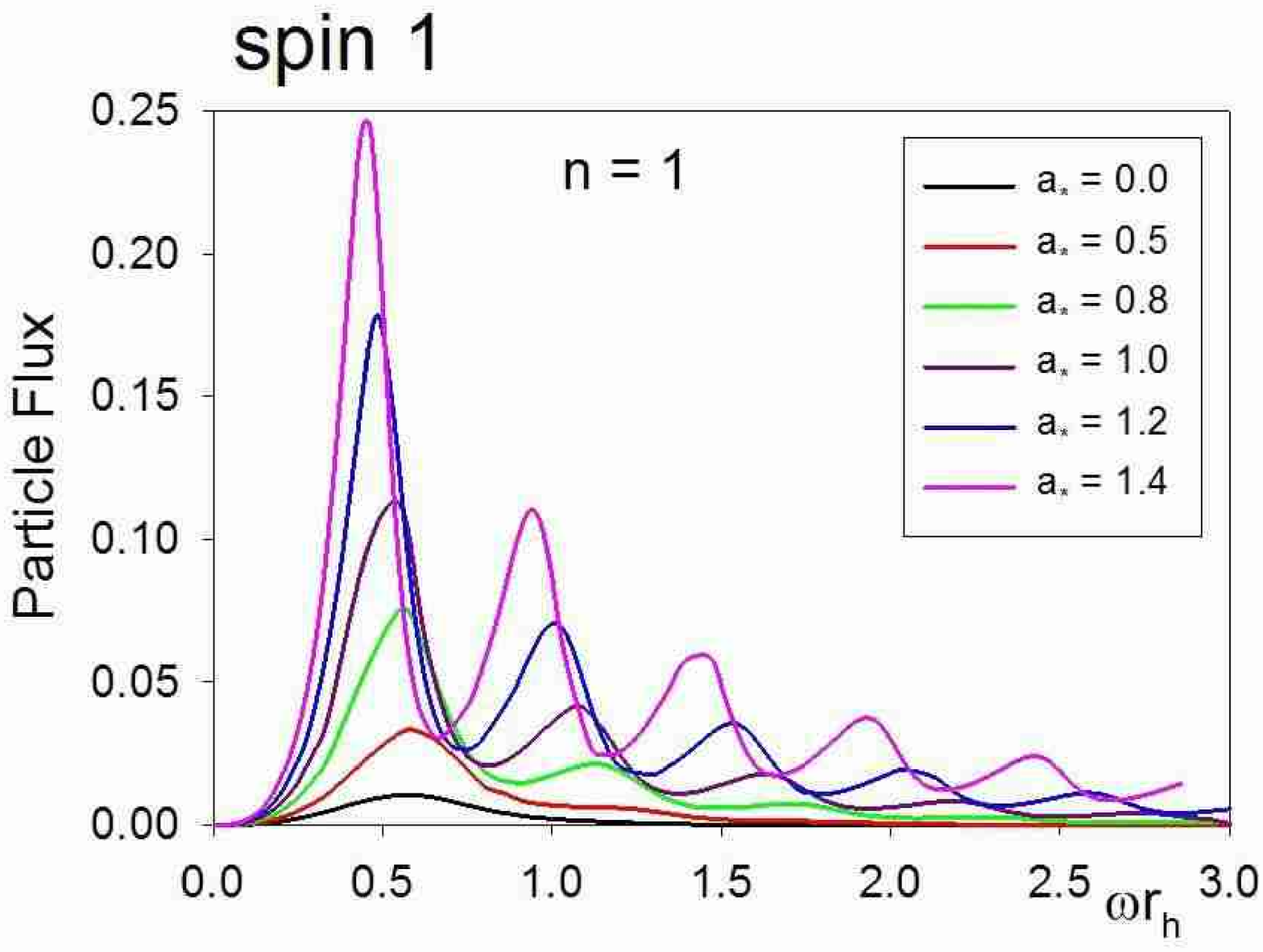}
\end{center}
\caption{Particle fluxes (\ref{eq:flux}) for spin-0, spin-1/2 and spin-1 fields
as a function of $\omega r_{h}$, for fixed $n=1$ and varying $a_{*}$.}
\label{fig:flux}
\end{figure}
\begin{figure}[p]
\begin{center}
\includegraphics[width=5cm,angle=270]{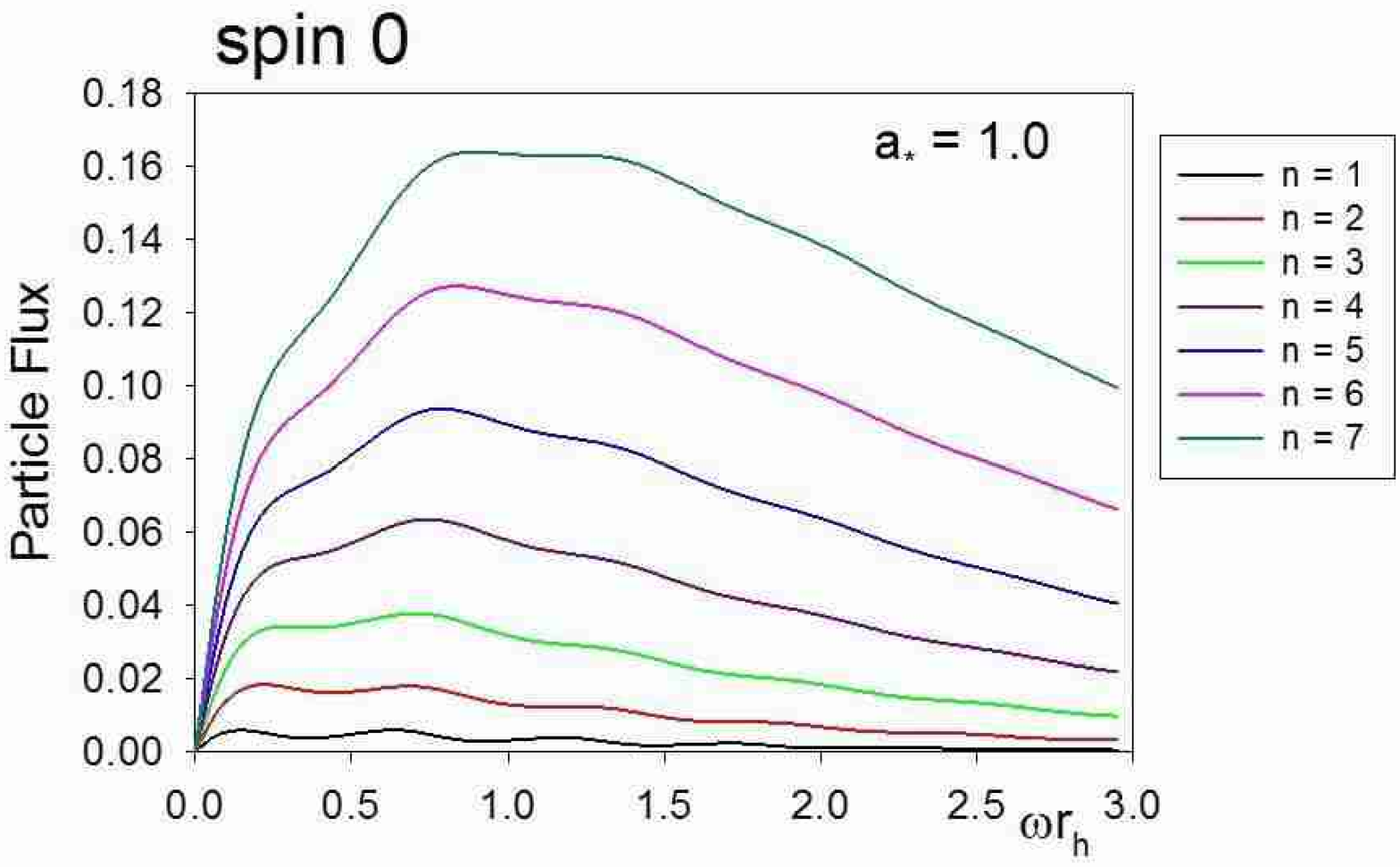}
\includegraphics[width=5cm,angle=270]{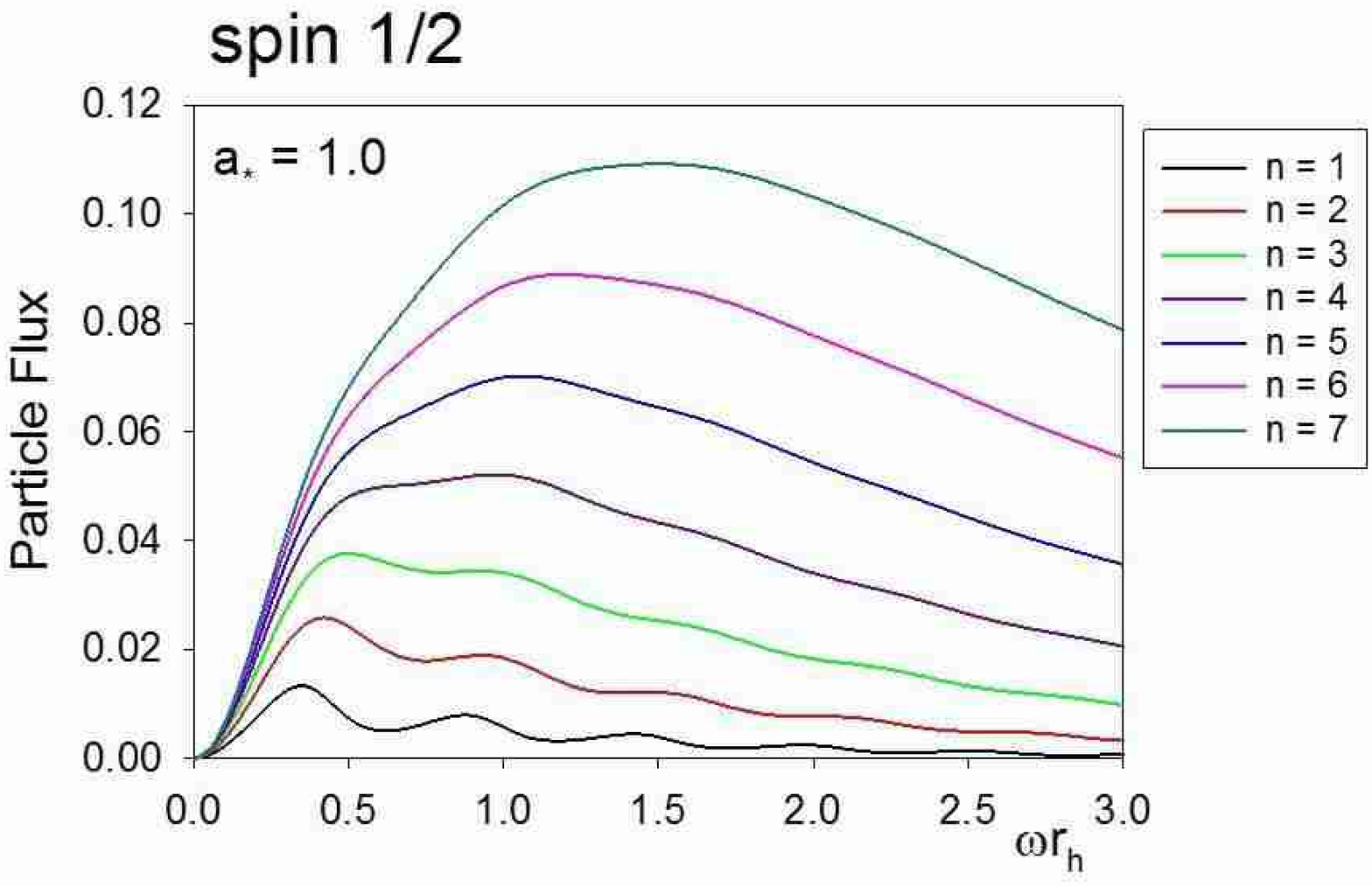}
\includegraphics[width=5cm,angle=270]{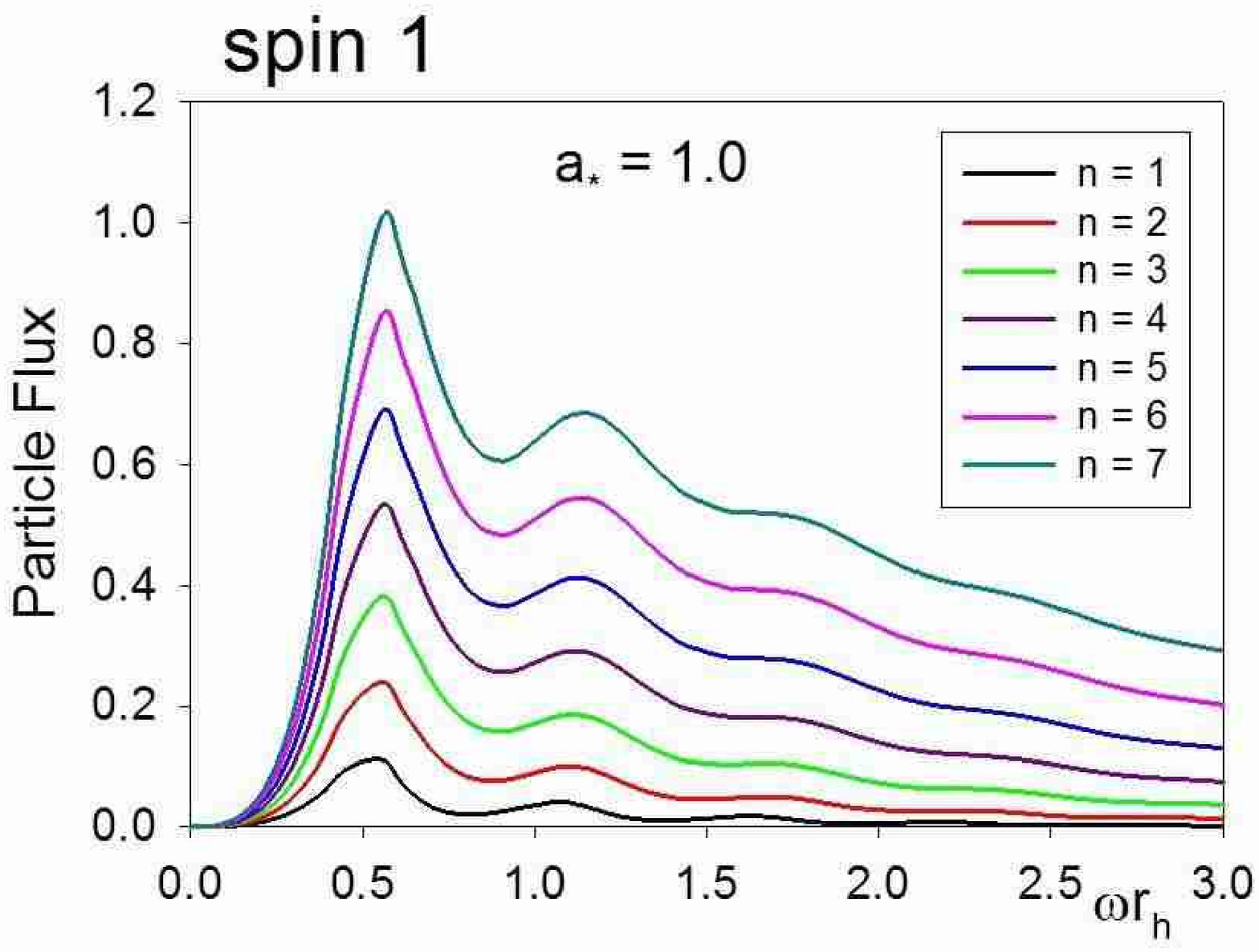}
\end{center}
\caption{Particle fluxes (\ref{eq:flux}) for spin-0, spin-1/2 and spin-1 fields as a function of $\omega r_{h}$,
for fixed $a_{*}=1$ and varying $n$.}
\label{fig:flux_a1}
\end{figure}

We now turn to the observable fluxes of particles, energy and angular momentum.
Firstly, the particle flux (\ref{eq:flux})
is shown in figures \ref{fig:flux} and \ref{fig:flux_a1} as a function of $\omega r_{h}$, for fixed $n=1$ and varying
$a_{*}$ (figure \ref{fig:flux}) and fixed $a_{*}=1$ and varying $n$ (figure \ref{fig:flux_a1}).

There are number of features that can be seen from figures \ref{fig:flux} and \ref{fig:flux_a1}:
\begin{enumerate}
\item
As the number of extra dimensions increases, so do the fluxes,
mostly because the Hawking temperature (\ref{temperature}) of the black hole increases linearly with $n$
for fixed $a_{*}$.
\item
For fixed $n$, as $a_{*}$ increases, the spectra become broader and flatter.
\item
For smaller values of $n$,  there are oscillations in the particle spectrum,
which correspond to various field modes becoming dominant at certain frequencies (a sort of resonance effect).
\item
Emission at higher frequencies becomes more important as $n$ or $a_{*}$ increase.
\item
The peaks for spin-0 and spin-1/2 particles are of the same order of magnitude,
but the peak in the emission for spin-1 particles is roughly ten times bigger.
\end{enumerate}

\begin{figure}[h]
\begin{center}
\includegraphics[width=6.5cm,angle=270]{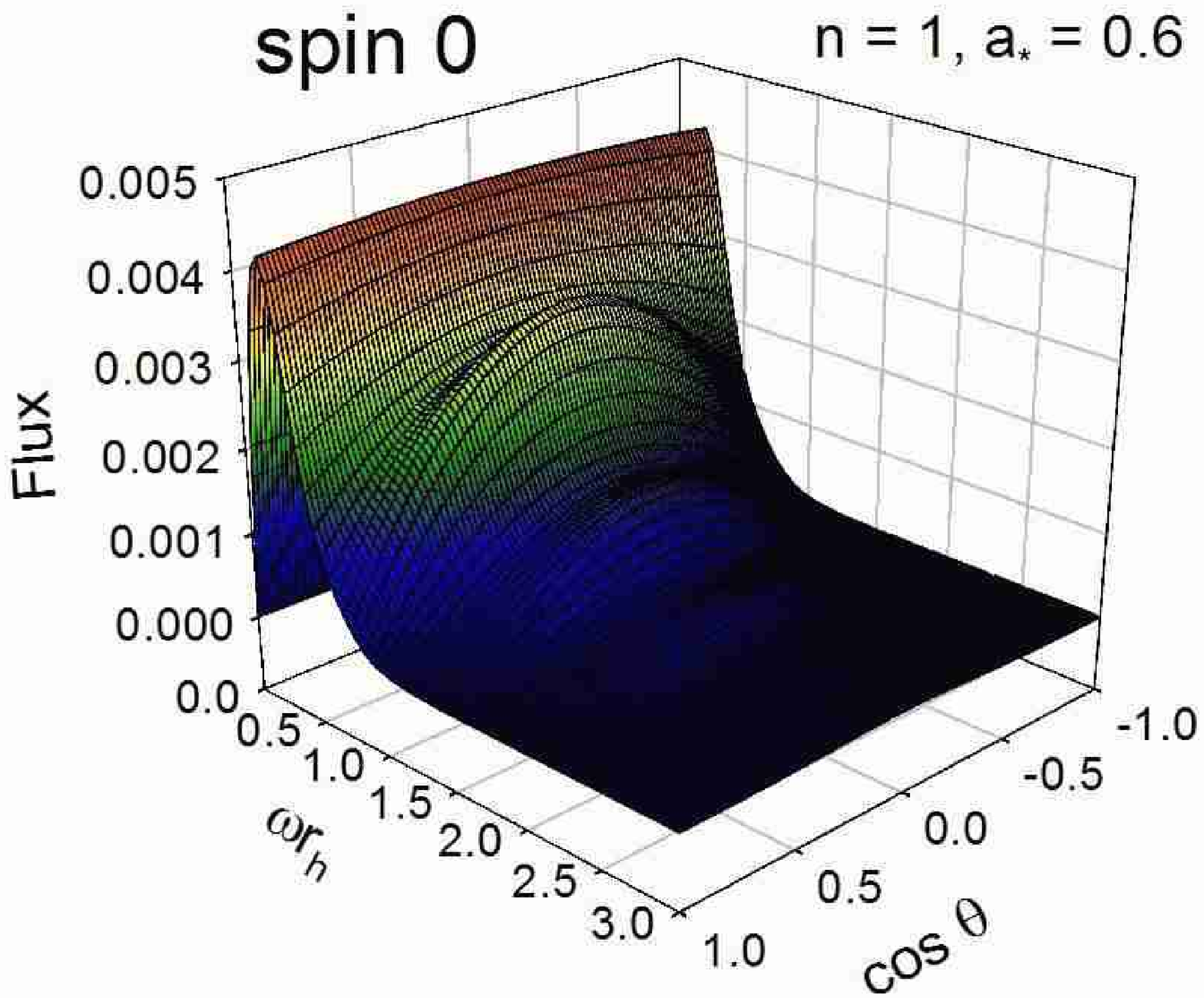}
\includegraphics[width=6.5cm,angle=270]{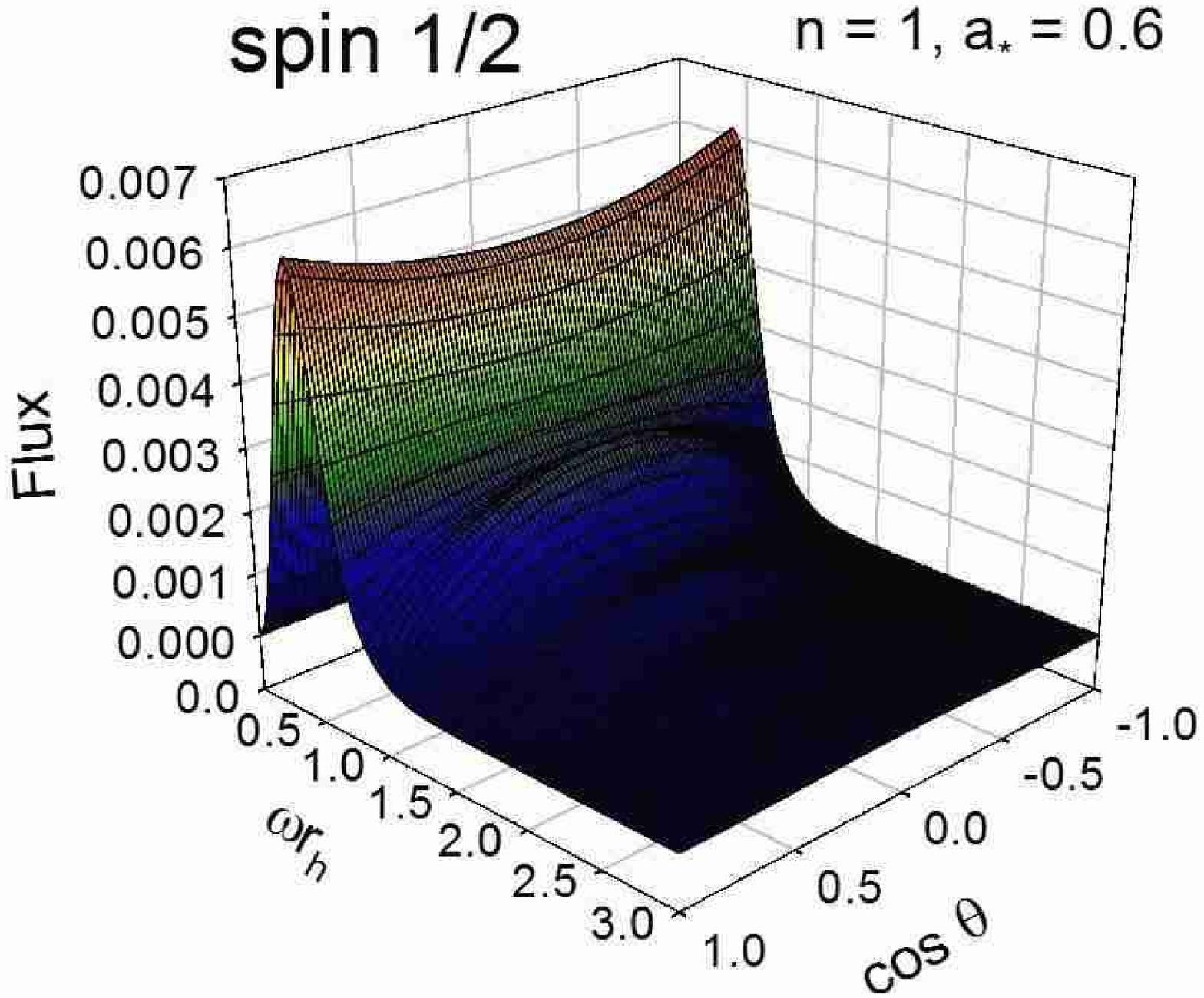}
\includegraphics[width=6.5cm,angle=270]{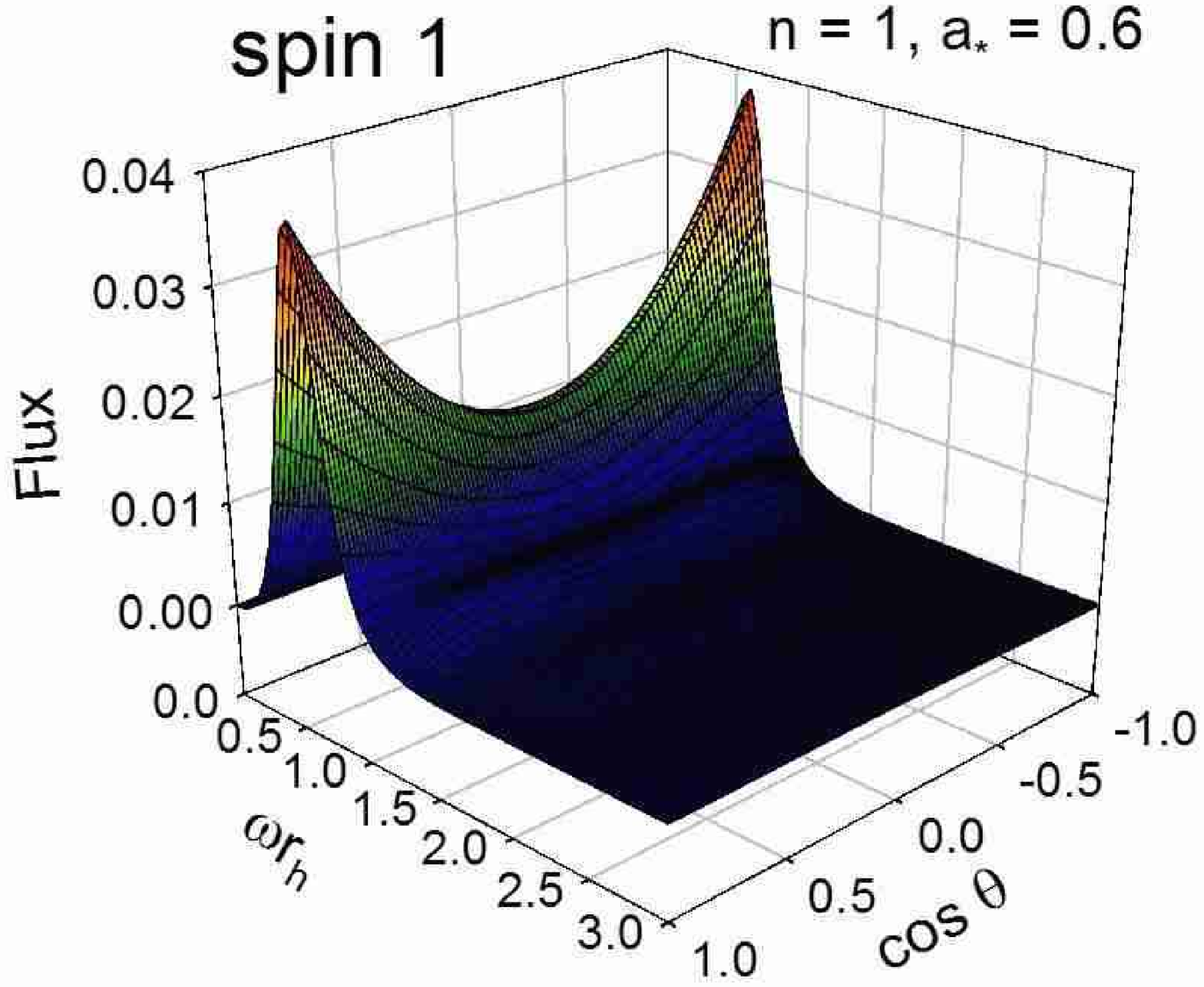}
\end{center}
\caption{Angular dependence of the particle fluxes (\ref{eq:fluxang}) for spin-0, spin-1/2 and spin-1 fields
for $n=1$ and $a_{*}=0.6$. Here, $\theta $ is the latitudinal angle, so that $\theta = 0,\pi $ corresponds
to the axis of rotation of the black hole and $\theta = \pi/2$ is the equatorial plane.}
\label{fig:fluxang}
\end{figure}

We have also studied how the particle flux depends on the latitudinal angle $\theta $ (\ref{eq:fluxang}),
and the results can be seen in figure \ref{fig:fluxang}.
In figure \ref{fig:fluxang}, we have $\cos \theta = \pm 1$ along the rotation axis of the black hole,
and $\cos \theta =0$ in the equatorial plane of the black hole.
It can be seen that the angular distributions are quite different for the three cases.
For the spin-0 case, we can again see the peaks in the particle flux.
For small $\omega $, the angular distribution is uniform and spherically symmetric.
However, for larger values of $\omega $,
the flux is concentrated in the equatorial plane of the black hole,
as might be expected due to the rotation of the black hole.
For the spin-1 case, the flux is again concentrated in the equatorial plane for large $\omega $,
but there is a more significant concentration of emission along the axis of the black hole,
for smaller values of $\omega $.
This difference in behaviour is due to the coupling of the rotation of the black hole with the
electromagnetic field (which has non-zero spin).
For the spin-1/2 case, the energy is again concentrated in the equatorial plane for large $\omega $,
and there is a very slight spin-orbit coupling effect for small values of $\omega $.

\begin{figure}[p]
\begin{center}
\includegraphics[width=4.5cm,angle=270]{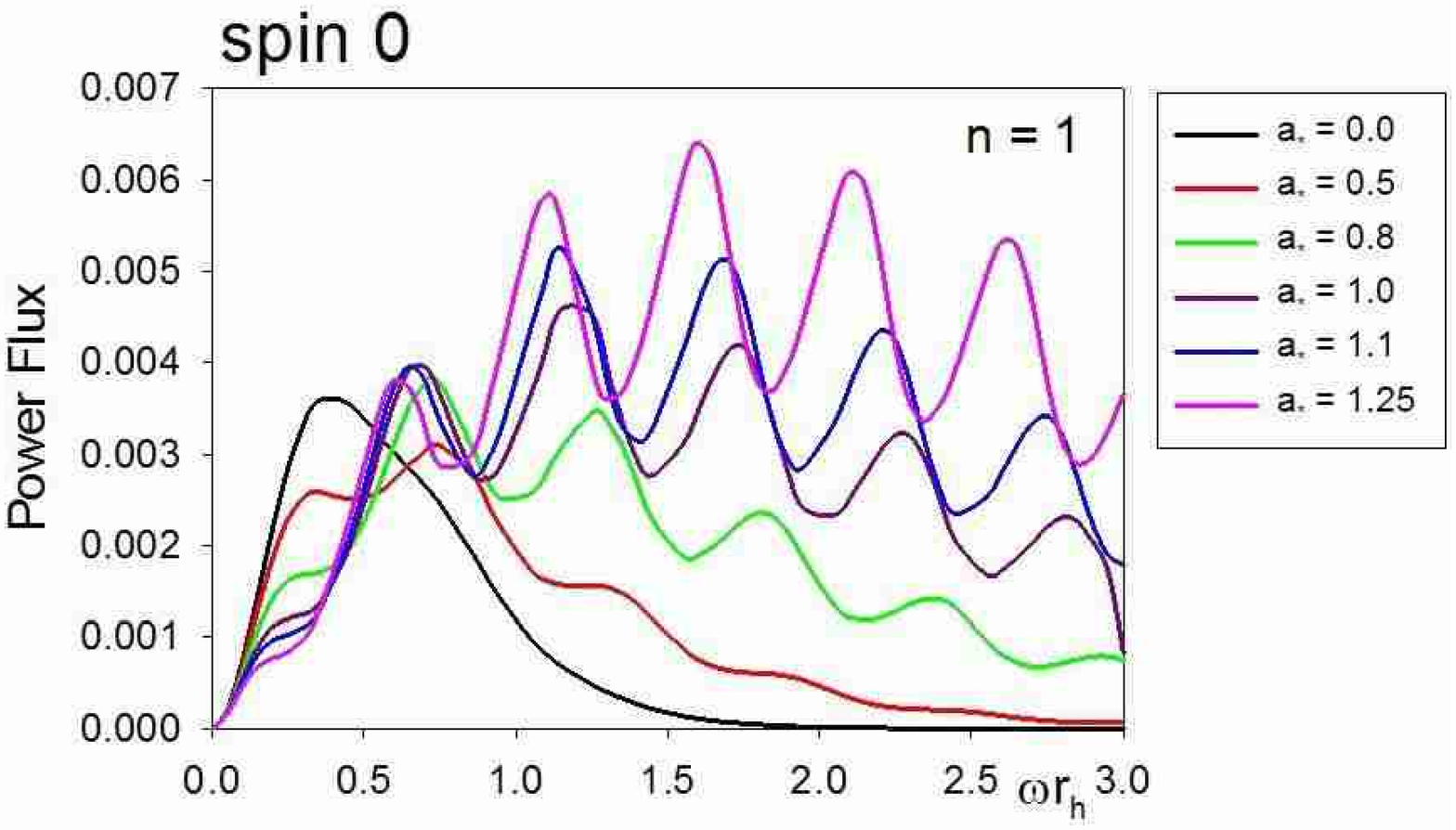}
\includegraphics[width=4.5cm,angle=270]{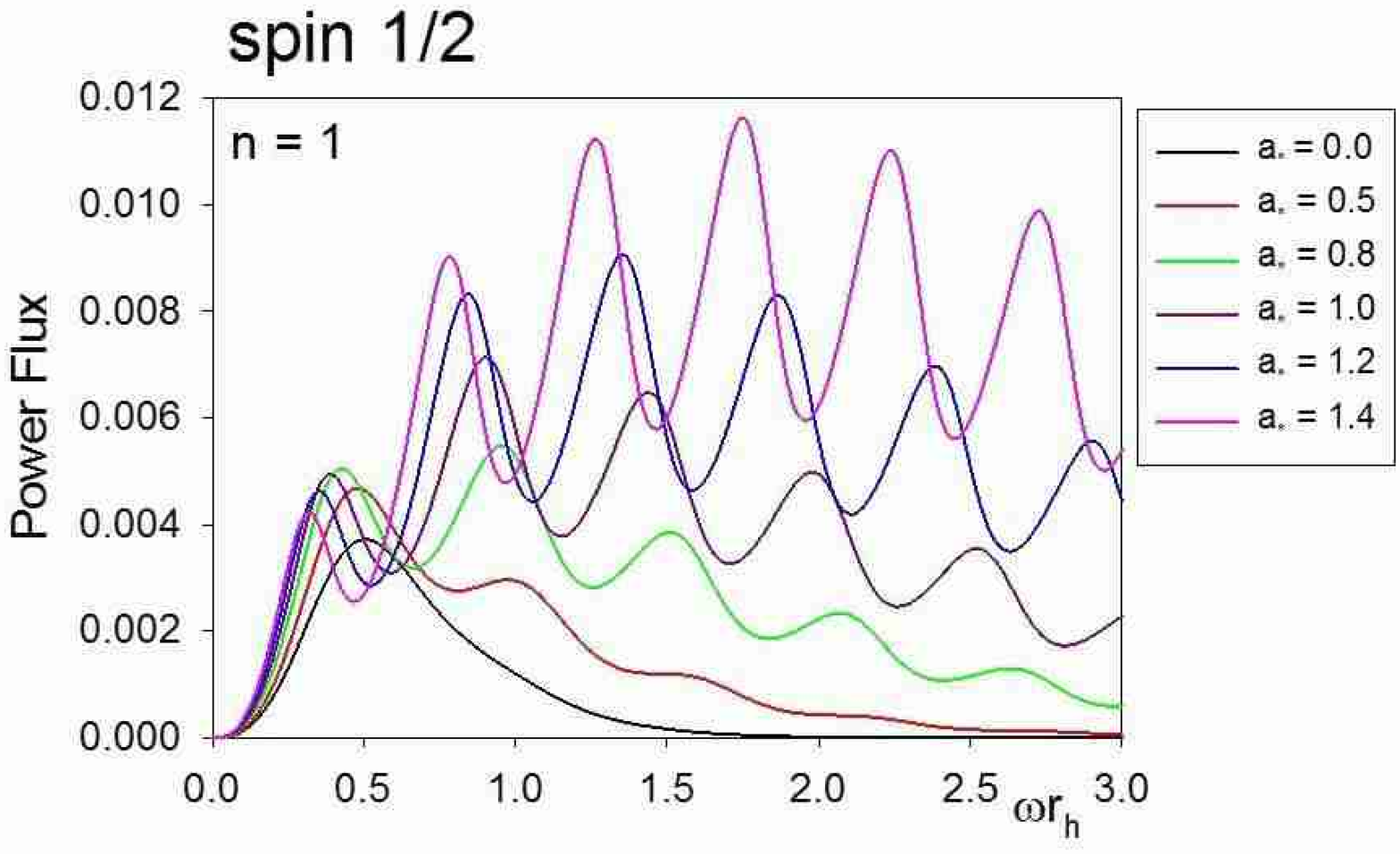}
\includegraphics[width=4.5cm,angle=270]{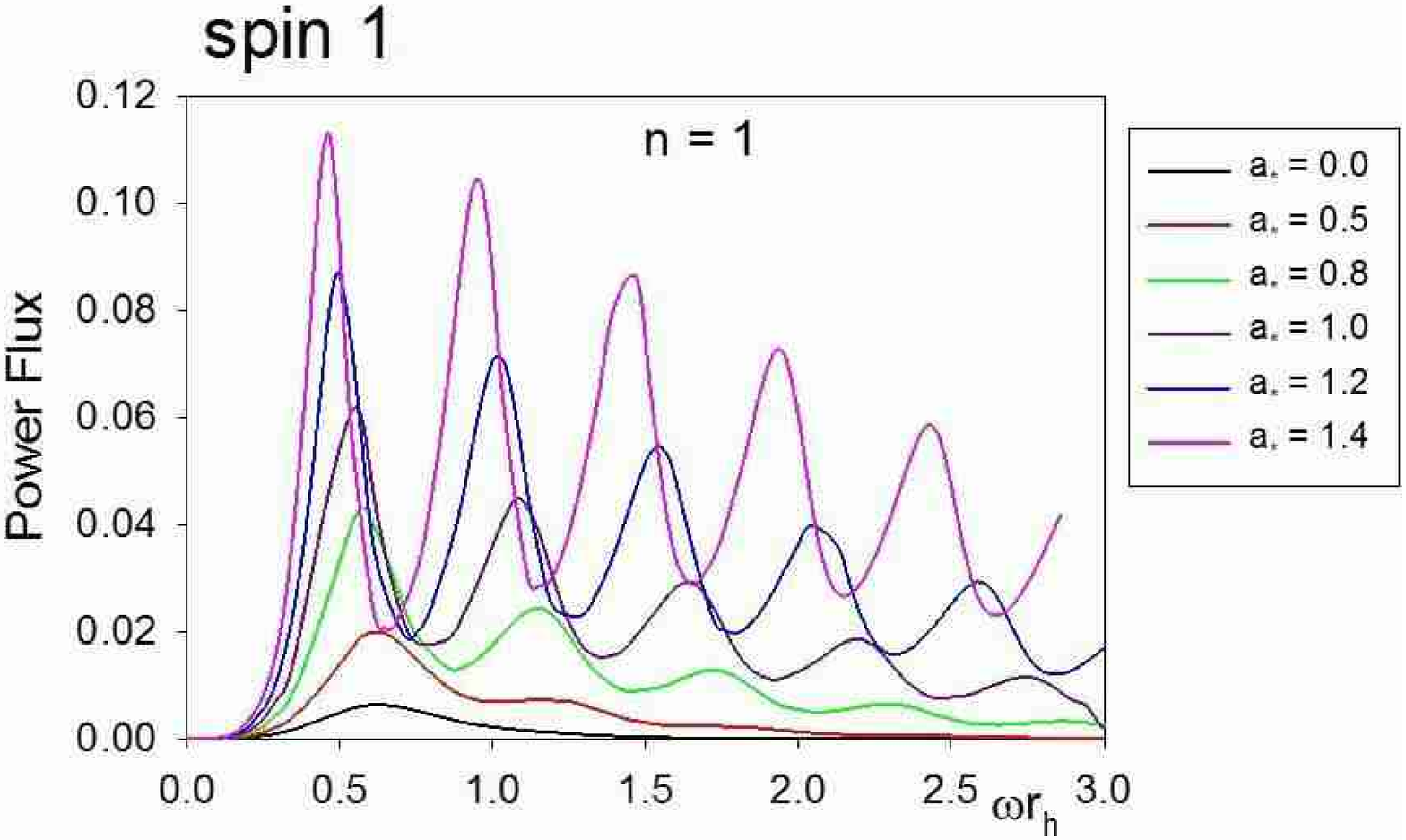}
\end{center}
\caption{Power fluxes (\ref{eq:power}) for spin-0, spin-1/2 and spin-1 fields as a function of $\omega r_{h}$, for fixed
$n=1$ and varying $a_{*}$.}
\label{fig:power}
\end{figure}
\begin{figure}[p]
\begin{center}
\includegraphics[width=5cm,angle=270]{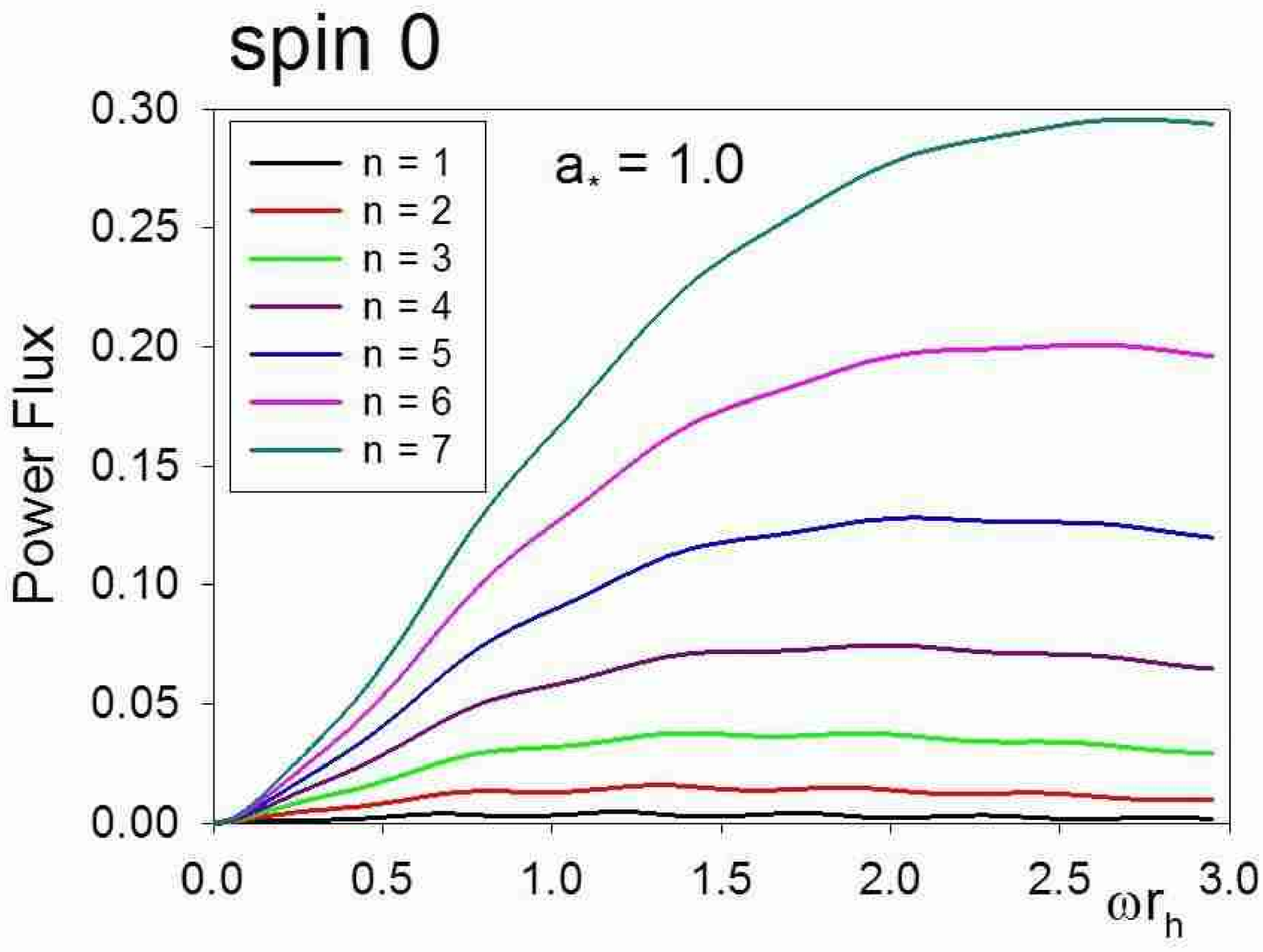}
\includegraphics[width=5cm,angle=270]{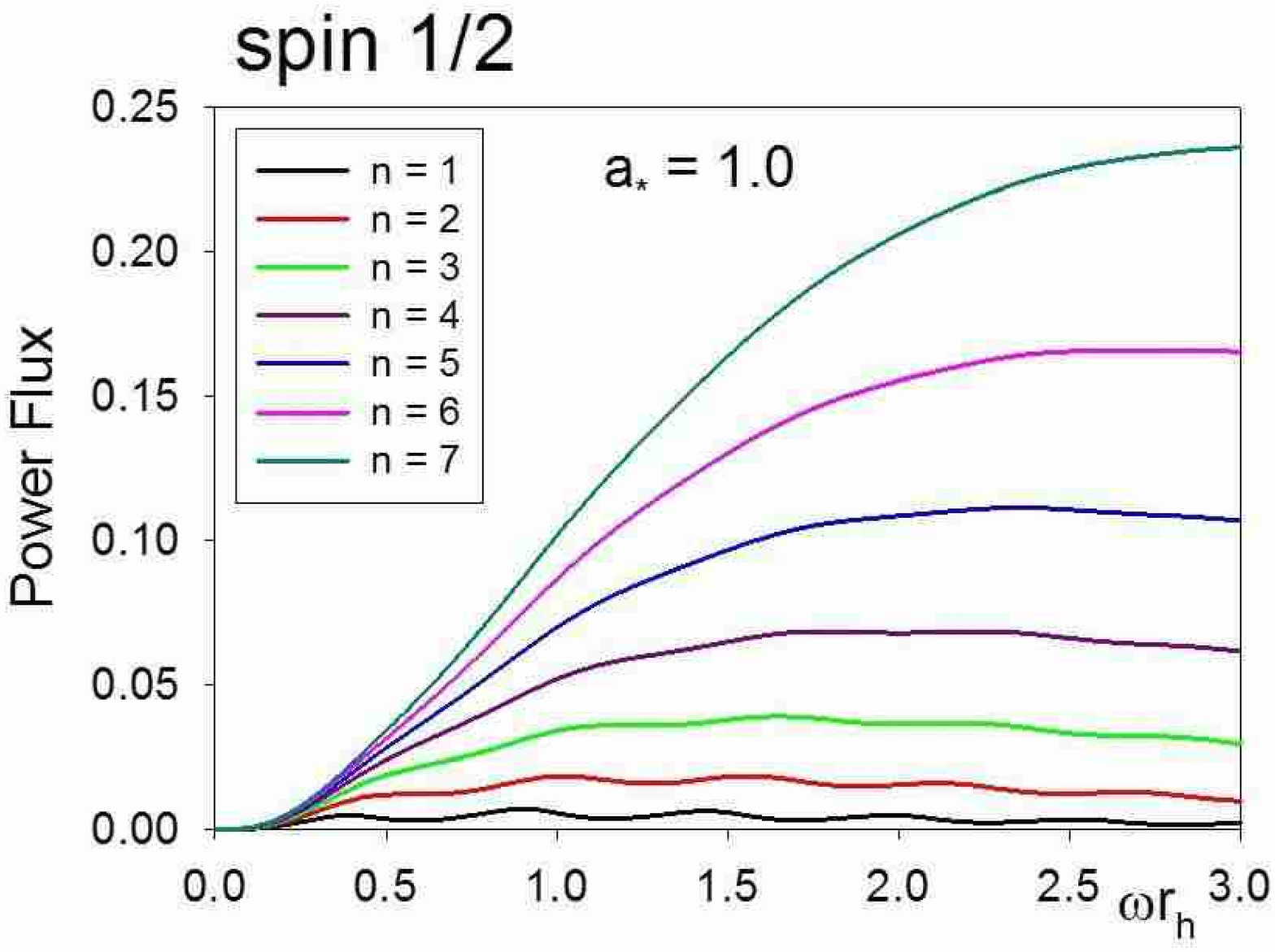}
\includegraphics[width=5cm,angle=270]{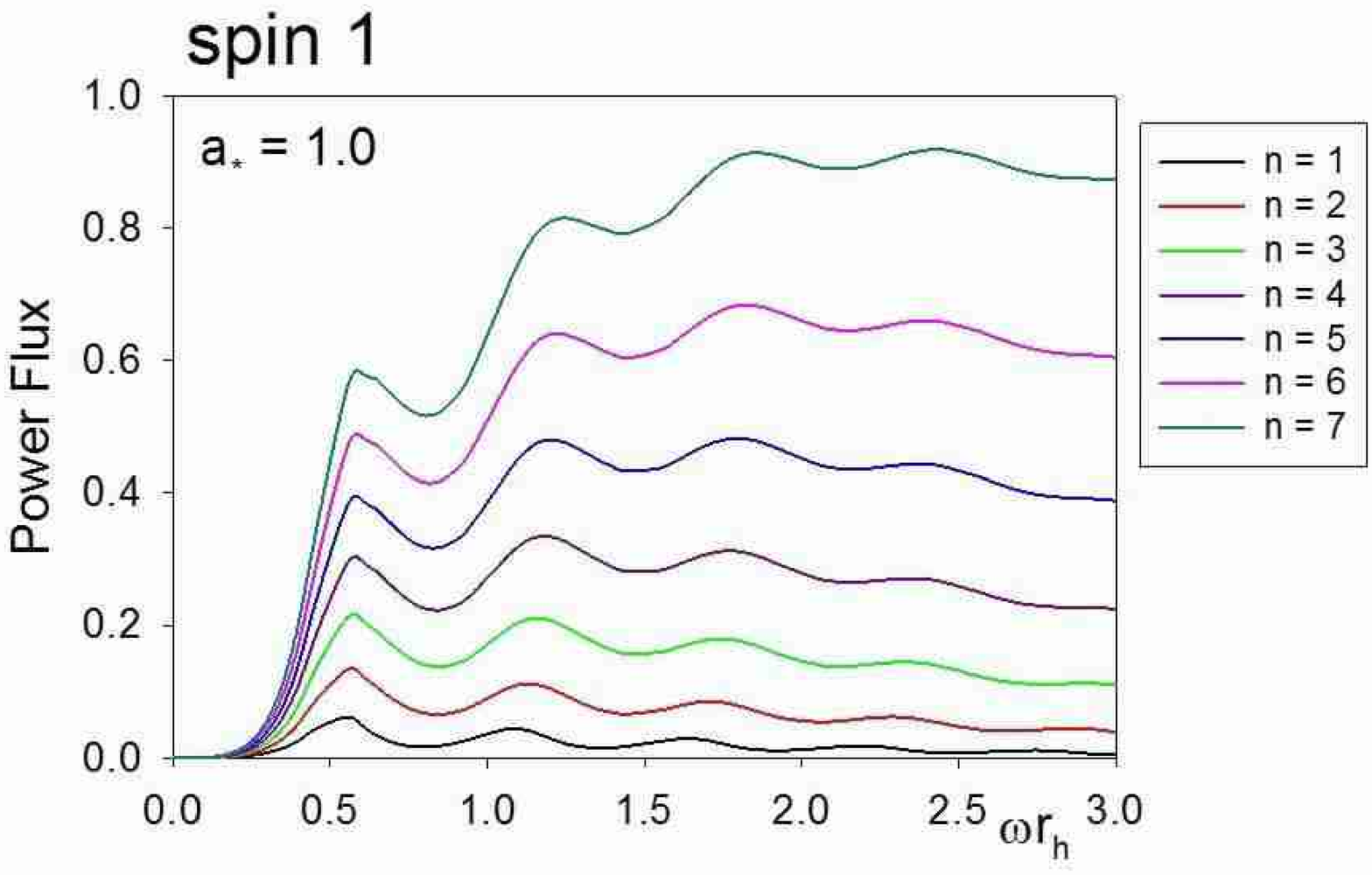}
\end{center}
\caption{Power fluxes (\ref{eq:power}) for spin-0, spin-1/2 and spin-1 fields as a function of $\omega r_{h}$, for fixed
$a_{*}=1$ and varying $n$.}
\label{fig:power_a1}
\end{figure}

Next, we consider the power flux (\ref{eq:power}), and plot in figures \ref{fig:power} and \ref{fig:power_a1}
the power fluxes as functions of $\omega r_{h}$, for fixed $n$ and fixed $a_{*}$, respectively.
Many of the features follow from those of the particle flux, as outlined above.
In particular, emission at higher values of $\omega $ becomes more important as either $n$ or $a_{*}$ increase.
This is enhanced compared with the particle flux because the power in each mode involves multiplying the particle
flux in that mode by $\omega $.

\begin{figure}[h]
\begin{center}
\includegraphics[width=6.5cm,angle=270]{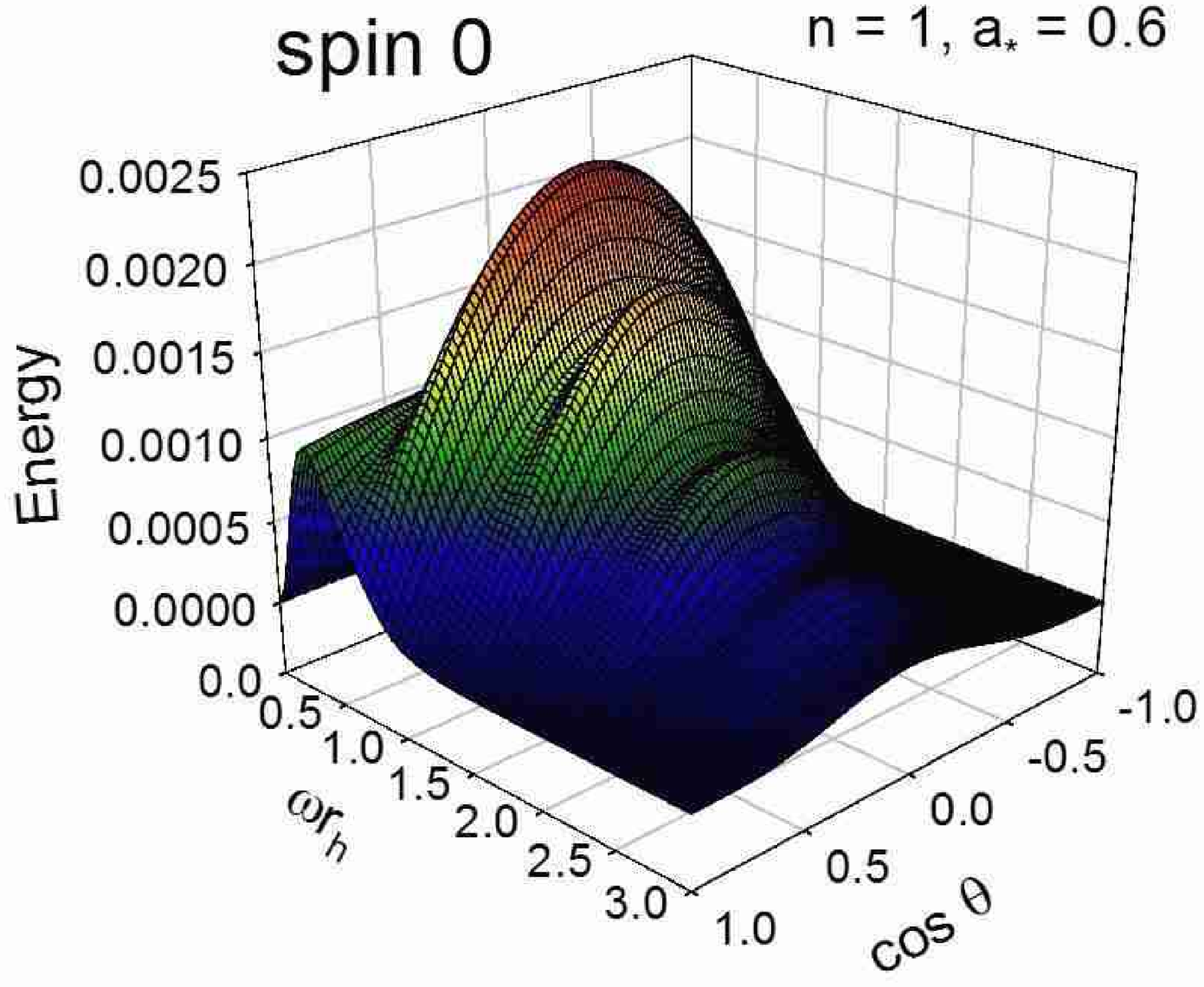}
\includegraphics[width=6.5cm,angle=270]{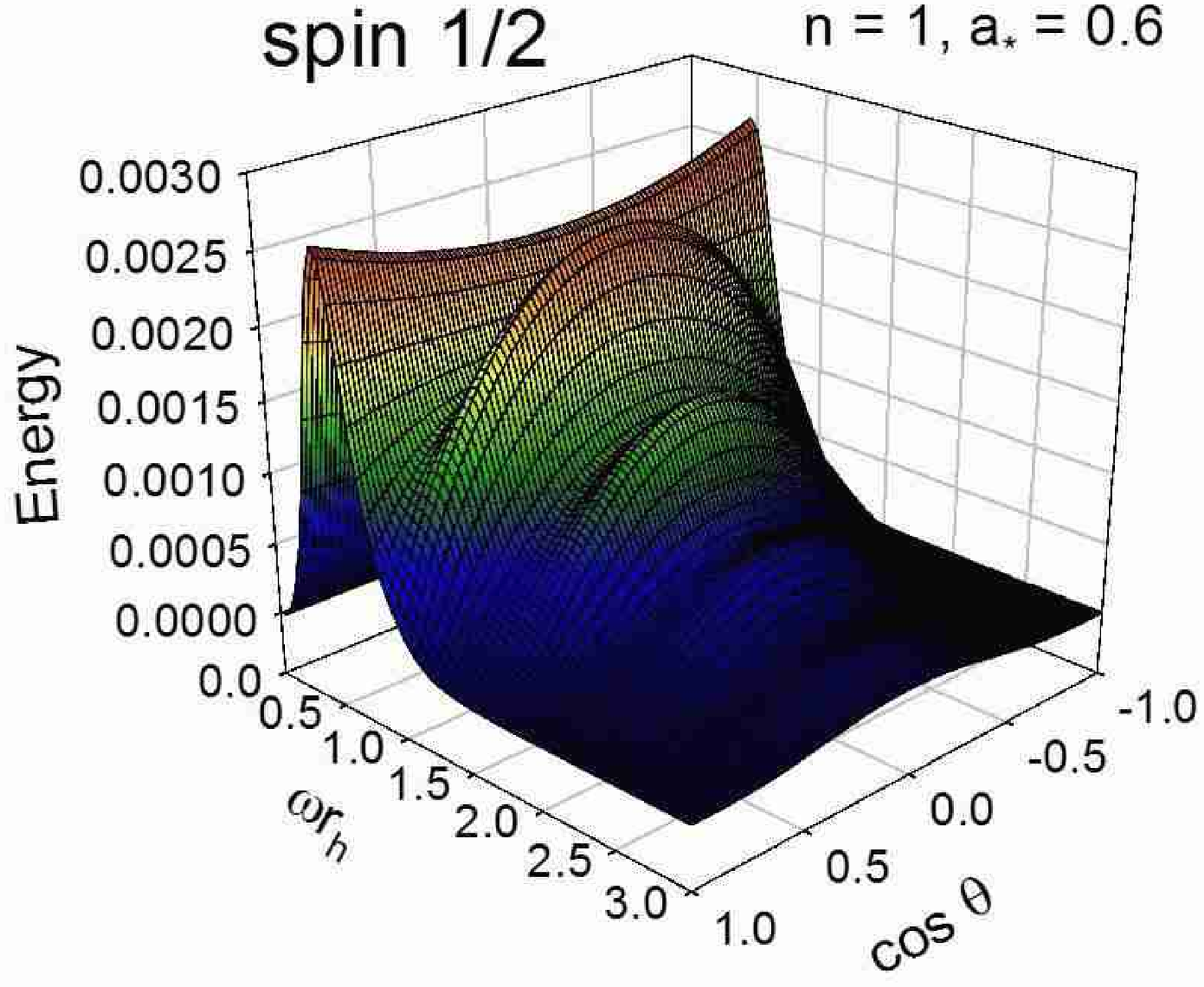}
\includegraphics[width=6.5cm,angle=270]{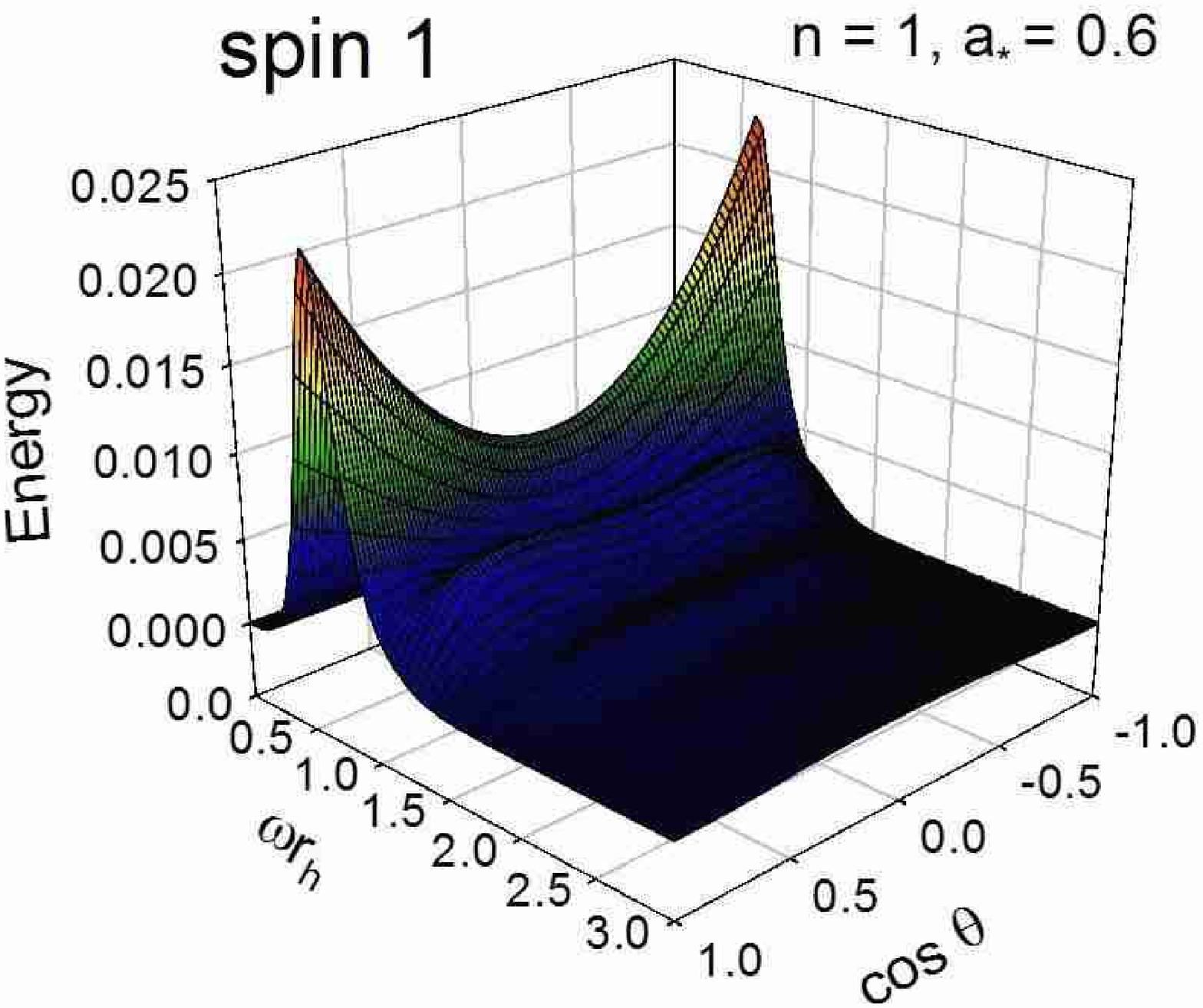}
\end{center}
\caption{Angular dependence of the energy fluxes (\ref{eq:powerang}) for spin-0, spin-1/2 and spin-1 fields
for $n=1$ and $a_{*}=0.6$. Here, $\theta $ is the latitudinal angle, so that $\theta = 0,\pi $ corresponds
to the axis of rotation of the black hole and $\theta = \pi/2$ is the equatorial plane.}
\label{fig:powerang}
\end{figure}

The angular dependence of the power flux  (\ref{eq:powerang})
is shown in figure \ref{fig:powerang} for $n=1$ and $a_{*}=0.6$.
The spin-orbit coupling is particularly clear at small values of $\omega $, and again the emission
is concentrated along the equatorial plane for large values of $\omega $.

\begin{figure}[p]
\begin{center}
\includegraphics[width=5cm,angle=270]{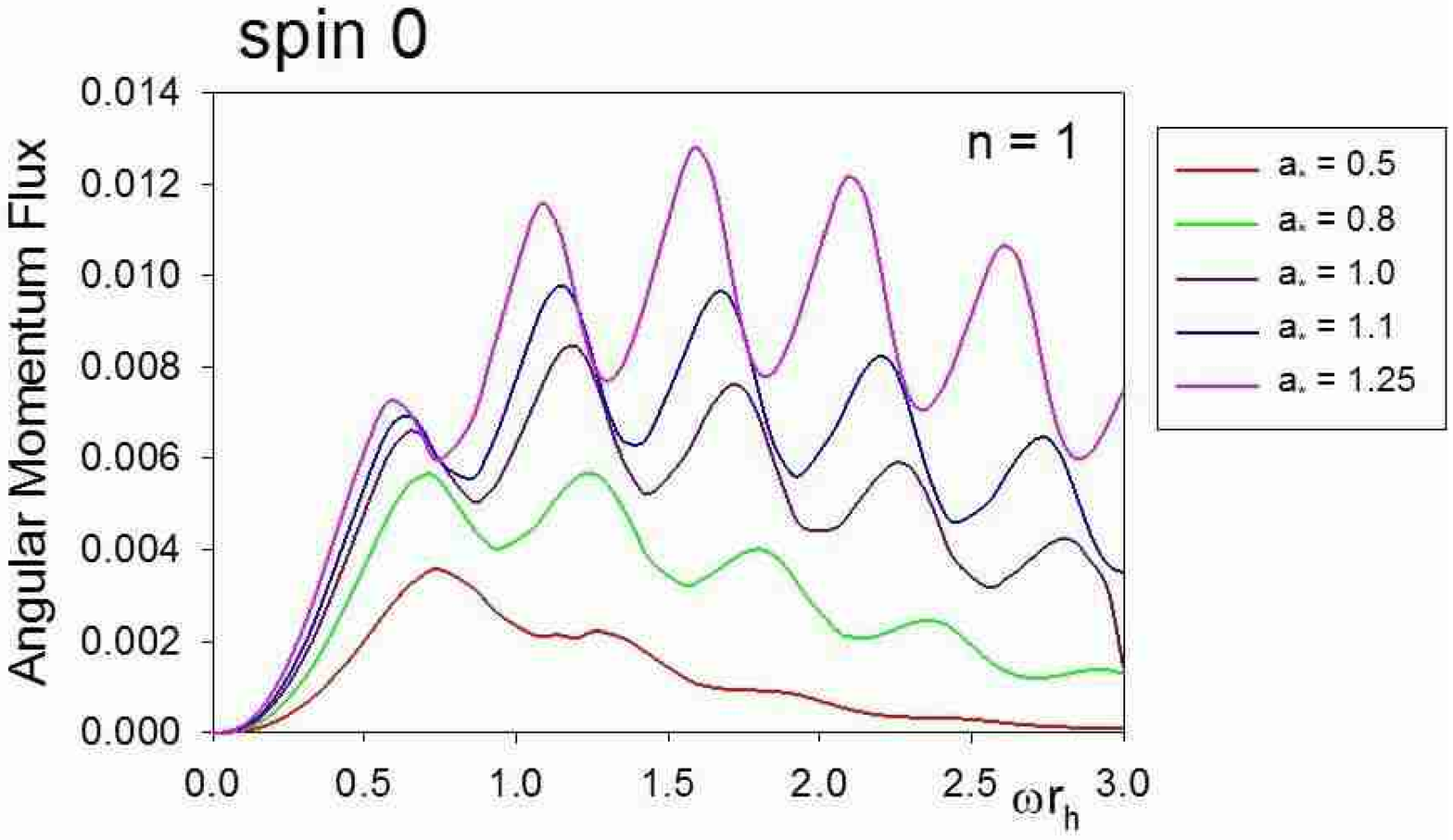}
\includegraphics[width=5cm,angle=270]{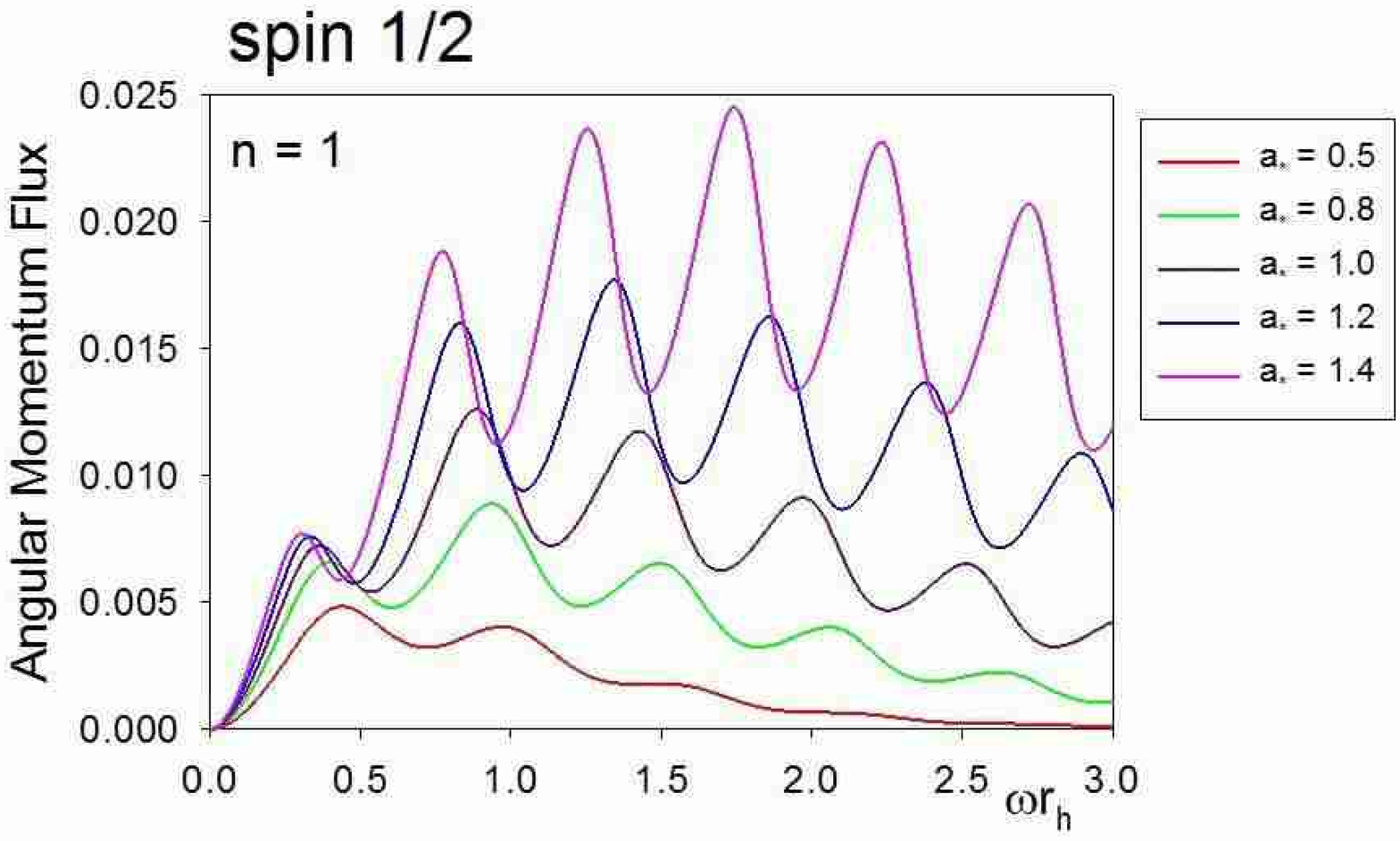}
\includegraphics[width=5cm,angle=270]{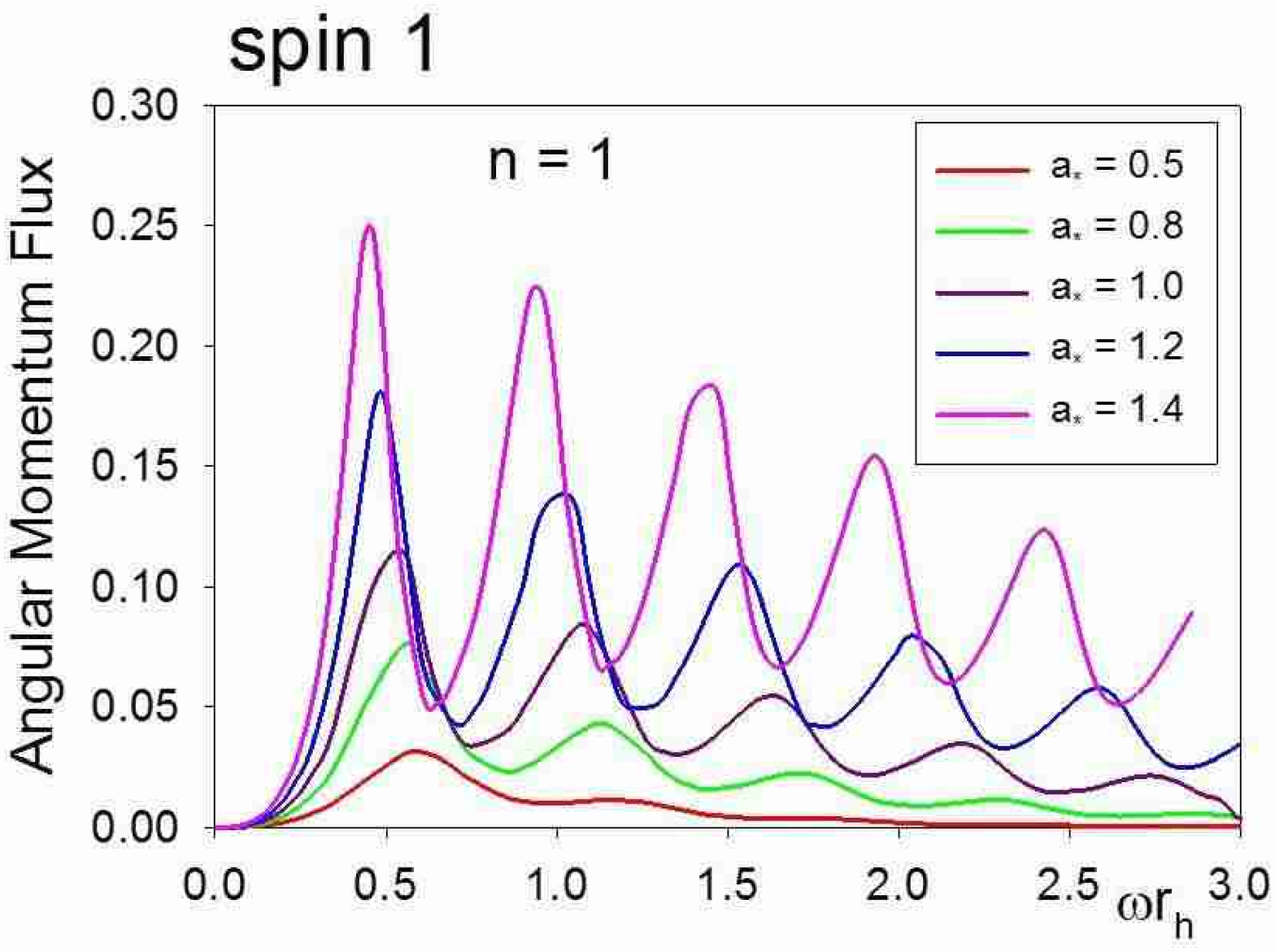}
\end{center}
\caption{Angular momentum fluxes (\ref{eq:angmom}) for spin-0, spin-1/2 and spin-1 as a function of $\omega r_{h}$,
for fixed $n=1$ and varying $a_{*}$.}
\label{fig:angmom}
\end{figure}
\begin{figure}[p]
\begin{center}
\includegraphics[width=5cm,angle=270]{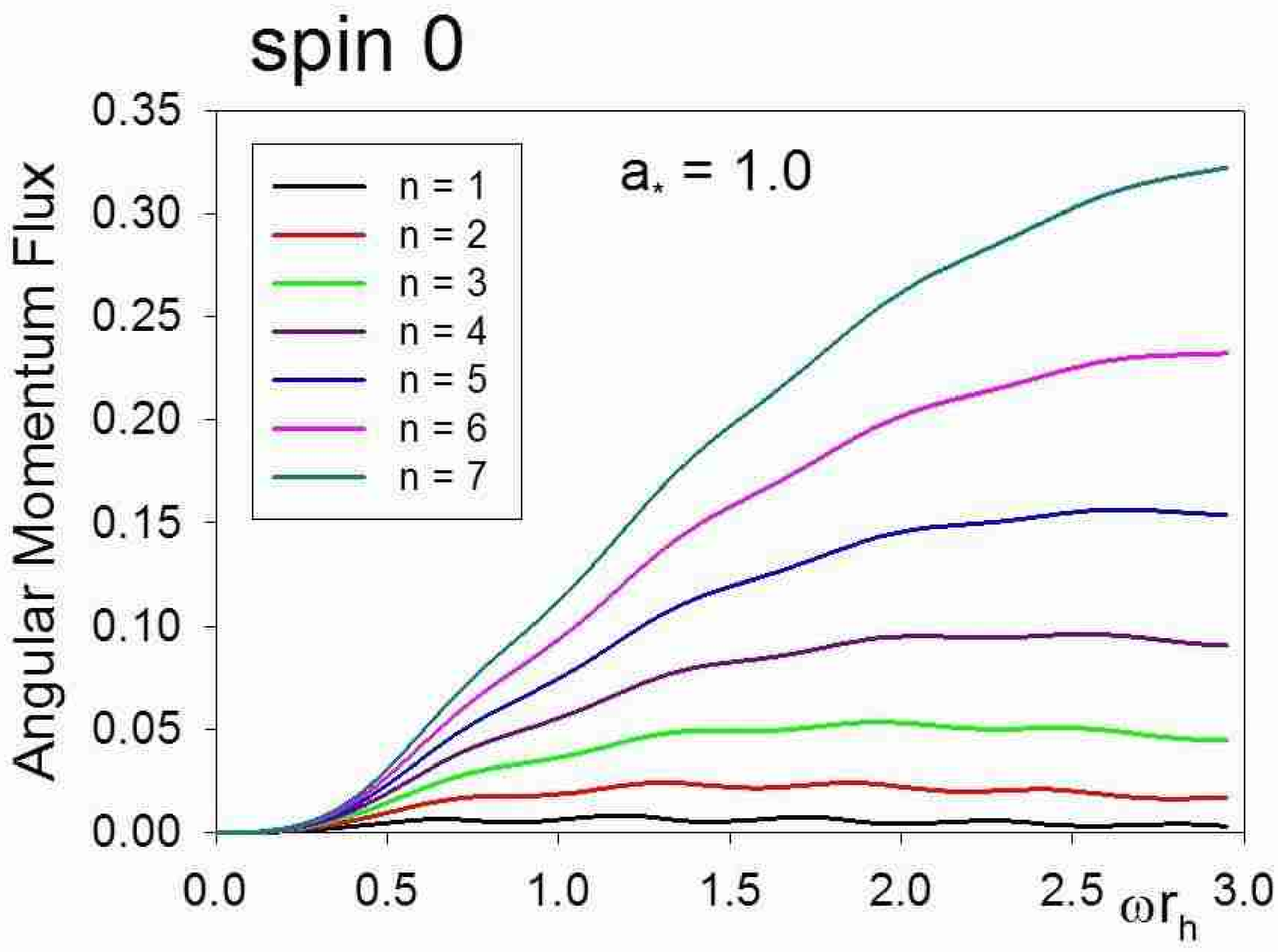}
\includegraphics[width=5cm,angle=270]{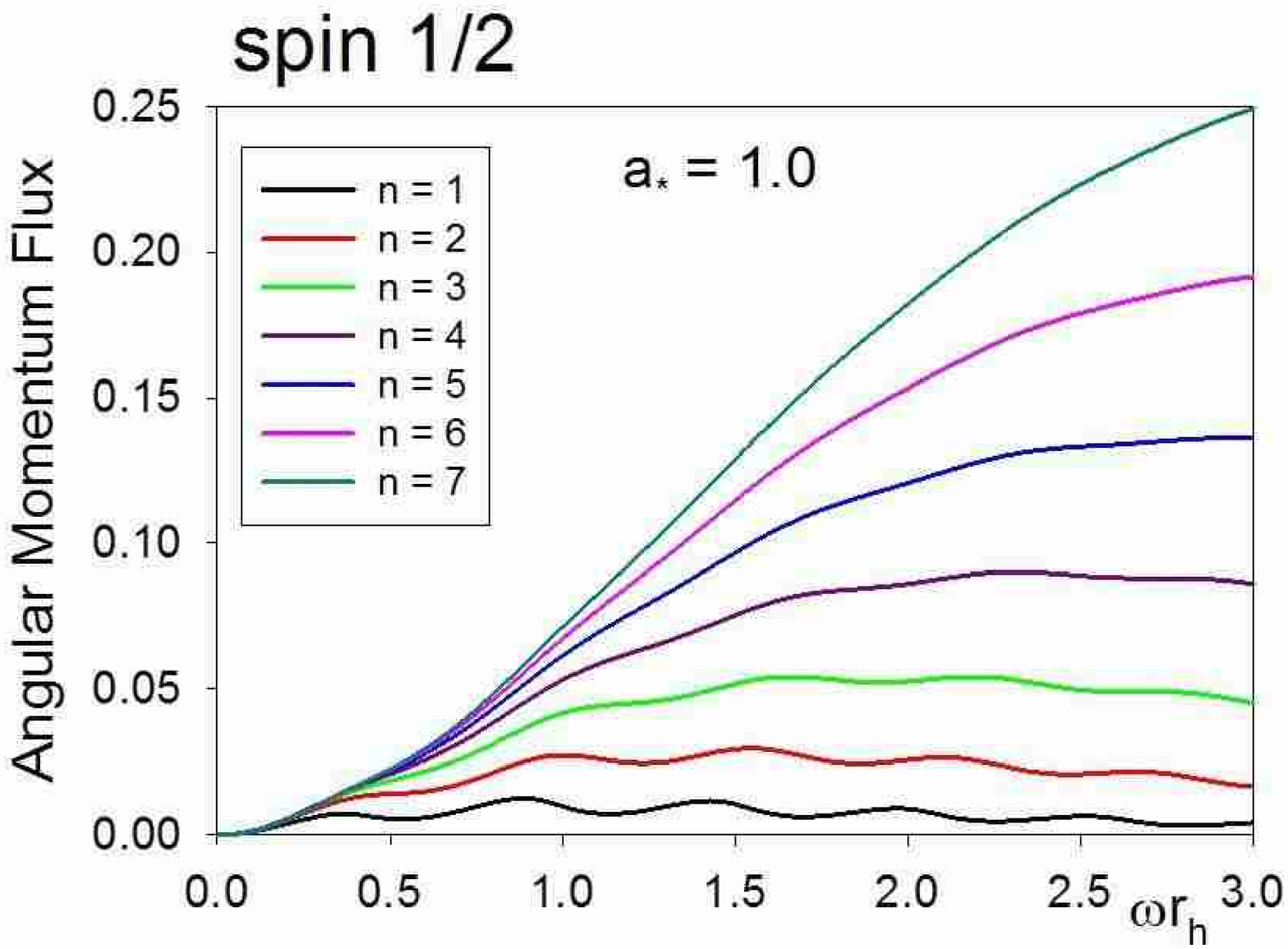}
\includegraphics[width=5cm,angle=270]{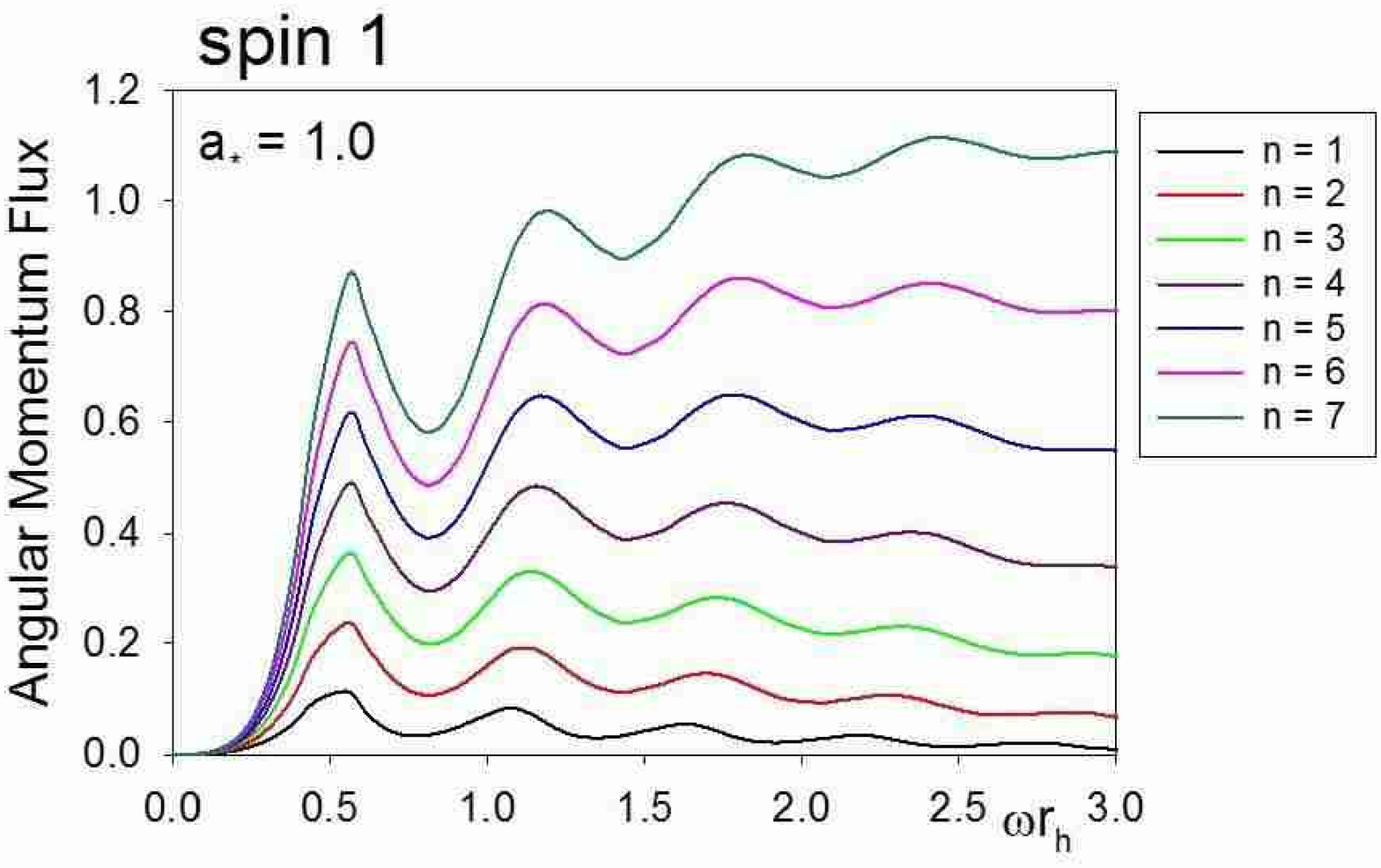}
\end{center}
\caption{Angular momentum fluxes (\ref{eq:angmom}) for spin-0, spin-1/2 and spin-1 fields
as a function of $\omega r_{h}$, for fixed $a_{*}=1$ and varying $n$.}
\label{fig:angmom_a1}
\end{figure}
Finally, we consider the rate at which the black hole loses angular momentum.
The flux of angular momentum (\ref{eq:angmom}) is shown in figures \ref{fig:angmom} and \ref{fig:angmom_a1}
for $n=1$ and $a_{*}=1$ respectively.
The main features are as would be anticipated from our previous analysis of the particle and energy fluxes.
The angular dependence of the flux of angular momentum has no experimental significance so we do not consider it here.

\section{Conclusions}
\label{sec:conc}

In this brief review we have considered the Hawking radiation of rotating brane black holes, as may be produced at
the LHC.
In references \cite{DHKW,CDKW,CKW}
we have studied in detail the fluxes of particles, energy and angular momentum for particles of spin-0, spin-1/2 and
spin-1, emitted on the brane.
This is sufficient to compute the emission of all standard model particles on the brane.
Our results will be of use for simulations of black hole events at the LHC.
There are currently two such simulation codes, CHARYBDIS \cite{Harris,charybdis} and CATFISH \cite{catfish}
(see \cite{Gingrich} for a recent comparison of these).

To date both simulators only use detailed information about the `Schwarzschild' phase of the black hole's evolution.
It is hoped that our detailed results on the `spin-down' phase will be incorporated into these simulators.
This will enable the effects of the rotation of the black hole on LHC events to be accurately modelled.
However, the analysis of the `spin-down' phase is not yet complete: information about
graviton emission (which will contribute `missing energy' to LHC events \cite{missing}) is required.
We hope to return to this question in the near future.

\section*{Acknowledgements}
I thank the organizers of the workshop on ``Dynamics and Thermodynamics
of Black Holes and Naked Singularities II'' for a most enjoyable and stimulating meeting.
The work in this talk was done jointly with the following collaborators:
Marc Casals, Sam Dolan and Gavin Duffy (all School of Mathematics, University College Dublin),
Chris Harris (University of Cambridge) and Panagiota Kanti (University of Durham).
This work is supported by UK PPARC, grant reference number PPA/G/S/2003/00082.

\end{document}